\documentclass[prd,showpacs,amsmath,amssymb,nofootinbib,superscriptaddress]{revtex4}
\usepackage{array,mathtools,amssymb,booktabs,multirow,diagbox,hhline}
\pdfoutput=1

\usepackage{graphicx}
\usepackage{dcolumn}
\usepackage{bm}
\usepackage{bbold}

\newcommand{\Od}{{\cal O}}
\newcommand{\tr}{\mbox{tr}}
\newcommand{\re}{\mbox{Re}\,}
\newcommand{\diag}{\mbox{diag}}

\newcommand{\intvol}{\int_0^{t_f} dt \int_{vol} d^3\vec{x}}
\newcommand{\mean}[1]{\left\langle{#1}\right\rangle}

\newcommand{\condl}{\mean{\bar q q}_l}

\newcommand{\IZ}{{\Bbb Z}}
\newcommand{\ID}{{\mathbb{1}}}
\newcommand{\lsim}{\raise.3ex\hbox{$<$\kern-.75em\lower1ex\hbox{$\sim$}}}

\begin{document}

\title{Chiral perturbation theory for nonzero chiral imbalance}
\author{D.Espriu}
\email{espriu@icc.ub.edu}
\affiliation{Departament of Quantum Physics and Astrophysics and Institut de Ci\'encies del Cosmos (ICCUB), Universitat de Barcelona, Mart\'i Franqu\'es 1, 08028 Barcelona, Spain.}
\author{A. G\'omez Nicola}
\email{gomez@fis.ucm.es}
\affiliation{Departamento de F\'{\i}sica Te\'orica and IPARCOS. Universidad Complutense de Madrid, Avenida Complutense s/n, 28040 Madrid, Spain}
\author{A. Vioque-Rodr\'iguez}
\email{avioque@ucm.es}
\affiliation{Departamento de F\'{\i}sica Te\'orica and IPARCOS. Universidad Complutense de Madrid, Avenida Complutense s/n, 28040 Madrid, Spain}

\begin{abstract}
We construct the most general low-energy effective lagrangian including local parity violating terms  parametrized by an axial chemical potential or chiral imbalance $\mu_5$, up to  $\Od(p^4)$ order in the chiral expansion  for two light flavours.  For that purpose, we work within the Chiral Perturbation Theory framework where only pseudo-NGB fields are included, following the external source method. The $\Od(p^2)$ lagrangian is only modified by constant terms, while the $\Od(p^4)$ one  includes new terms proportional to $\mu_5^2$ and new low-energy constants (LEC), which are renormalized and related to particular observables. In particular, we analyze the corrections to the pion dispersion relation and observables related to the vacuum energy density, namely the light quark condensate, the chiral and topological susceptibilities and  the chiral charge density, providing numerical determinations of the new LEC when possible.  In particular,  we explore the dependence of the chiral restoration temperature $T_c$ with $\mu_5$. An increasing $T_c(\mu_5)$ is consistent with our fits to lattice data of the ChPT-based expressions. Although lattice uncertainties are still large and translate into the new LEC determination, a consistent physical description of  those observables emerges from our present work, providing a theoretically robust model-independent framework for further study of physical systems where parity-breaking effects may be relevant, such as heavy-ion collisions.
\end{abstract}

\maketitle

\section{Introduction}
\label{sec:intro} 

The possibility of the existence of space-time regions where parity is violated locally in QCD has attracted a lot of attention over recent years, mostly motivated by appealing theoretical proposals such as the Chiral Magnetic Effect (CME) \cite{Fukushima:2008xe,Kharzeev:2013ffa,Fukushima:2012fg}. Thus, local metastable $P$-breaking configurations can be created out of the QCD vacuum, still preserving global $P$ conservation, giving rise to observable effects when coupled to the magnetic field created in heavy-ion collisions. The same effect can  lead to interesting applications in condensed matter physics \cite{Kharzeev:2013ffa}.  The presence of such local $P$-breaking configurations can influence other observables in heavy-ion collisions, such as the dilepton spectrum  \cite{Andrianov:2012hq,Andrianov:2014uoa}. 

A convenient way to parametrize such a $P$-breaking source or chiral imbalance is by means of a constant axial chemical potential $\mu_5$ to be added to the QCD action over a given finite space-time region. The axial current is not conserved at the quantum level due to the $U(1)_A$ axial anomaly equation. However, it is conserved at the lagrangian level in the massless limit. Thus, from  
the Atiyah-Singer index theorem, the chiral charge

\begin{equation}
Q_5=\int_{vol} d^3\vec{x} J^0_{5}(x)
\label{chiralcharge}
\end{equation}
satisfies $\mean{Q_5}=N_L-N_R$ with $N_{L,R}$ the number of left (right) zero modes of the Dirac operator. The characteristic time of $L-R$ quark oscillations is of order $1/m_q$ \cite{Andrianov:2017meh} which is much larger than the typical fireball duration at least for $m_{u,d}$. This supports that for the light $u,d$ quarks $Q_5$ may remain approximately conserved during the fireball evolution in a typical heavy-ion collision, giving rise in the light quark sector to a chemical potential term even for nonzero light quark masses: 

\begin{equation}
\intvol \ {\cal L}_{QCD}^0\rightarrow \intvol \ {\cal L}_{QCD}^0+\mu_5\intvol J^0_5(x)
\label{mu5rep}
\end{equation}

The previous replacement is equivalent to consider an axial source 

\begin{equation}
a_\mu^0=\mu_5 \delta_{\mu 0} 
\label{amu5}
\end{equation}
in the QCD generating functional $Z_{QCD}\left[v,a,s,p,\theta\right]$ in the presence of vector, axial, scalar, pseudoscalar and $\theta$ sources \cite{Gasser:1983yg,Gasser:1984gg}. Equivalently, one can perform a $U(1)_A$ rotation on the quark fields $q\rightarrow q'= \exp\left[i \beta(x) \gamma_5\right] q$ and choose $\beta(x)=\frac{\theta (x)}{2N_f}$, which allows to trade the axial and $\theta$ terms in the absence of additional vector or axial sources:

\begin{equation}
Z_{QCD}\left[0,0,{\cal M},0,\theta(x)\right]=Z_{QCD}\left[0,\frac{1}{2N_f}\partial_\mu \theta (x)\ID,{\cal M}\cos\left[\theta(x)/N_f\right],{\cal M}\sin\left[\theta(x)/N_f \right],0\right]
\label{zqcdtransf1}
\end{equation}
where ${\cal M}$ is the quark mass matrix. Thus, in the chiral limit ${\cal M}=0$, the chemical potential term in \eqref{mu5rep} is equivalent to a non-constant $\theta$ source $\theta(t)=\mu_5t+\theta_0$. It is also equivalent to a chemical potential related to the Chern-Simons topological current \cite{Andrianov:2017meh}.

In this context, it is important to provide theoretical support for the behaviour of QCD and hadronic observables in the presence of chiral imbalanced matter, specially  regarding the finite-temperature and finite-volume dependence around the QCD phase diagram, given its importance for heavy-ion  collisions. 

Different models have been considered recently to address this problem, including the Polyakov loop Linear Sigma Model \cite{Chernodub:2011fr}, NJL-like models \cite{Fukushima:2010fe,Gatto:2011wc,Andrianov:2013dta,Yu:2015hym,Braguta:2016aov,Ruggieri:2016ejz,Khunjua:2019lbv,Khunjua:2019ini} and a generalized sigma model including all the members of the scalar/pseudoscalar multiplets of isospin $I=0,1$ and the $\eta'$ \cite{Andrianov:2012dj}.  In those works, several relevant properties have been discussed, such as phase diagram features, the topological susceptibilty, the chiral density, the quark condensate and the meson dispersion relation. However, the results are not fully in agreement between different models, as we discuss below, in particular regarding the behaviour of chiral symmetry restoration with increasing axial chemical potential.

On the other hand, there have been a few analyses trying to reproduce such parity-breaking effects in the lattice. Although the pion masses used so far are still large and the continuum extrapolation is not entirely understood, there are promising results which may help to disentangle between model predictions if more precision is achieved in the near future. Thus, in  \cite{Yamamoto:2011ks} the chiral charge density and the CME  have been  investigated, while   the  dependence of the chiral restoration temperature with $\mu_5$ has been studied through the chiral condensate, scalar susceptibility and Polyakov loop for $N_c=2$ \cite{Braguta:2015zta} and $N_c=3$ \cite{Braguta:2015owi}. The chiral anomaly in the lattice has been also studied in \cite{Feng:2017dom} while in more recent lattice analysis, updated results for $N_f=2$ are provided on the chiral charge density, as well as the topological susceptibility and charge \cite{Astrakhantsev:2019wnp}.  

One of the puzzles which is still not fully understood is that the lattice results  clearly show growing condensate and $T_c$ with $\mu_5$ \cite{Braguta:2015zta,Braguta:2015owi} while theoretical analyses yield contradictory results. Thus,  within the NJL  framework, some works \cite{Fukushima:2010fe, Chernodub:2011fr,Gatto:2011wc} found the opposite behaviour, i.e. a decreasing $T_c(\mu_5)$ wile others \cite{Braguta:2016aov,Ruggieri:2016ejz} agree with the lattice results. This contradiction seems to be related to the choice of the regularization scheme  \cite{Yu:2015hym,Farias:2016let}. In addition, the sigma model approach in \cite{Chernodub:2011fr} is also in disagreement with  the lattice, while an analysis based on Schwinger-Dyson equations gives rise also to an  increasing $T_c(\mu_5)$ behaviour \cite{Xu:2015vna}. The general arguments given in \cite{Hanada:2011jb} as well as the growing of the constituent mass with $\mu_5$ found in \cite{Andrianov:2013dta} 
support also a growing quark condensate.

Our purpose here is to provide a model-independent approach, aiming to construct the most general effective lagrangian for the lightest degrees of freedom in the presence of the $\mu_5$ source. Preliminary ideas along this line have been proposed in \cite{Andrianov:2019fwz}.  Although this requires by definition that the applicability range is restricted to low $\mu_5$ and low temperatures, which poses certain limits e.g. on chiral restoration, our analysis will serve as a guideline for  models and lattice analyses, which should satisfy the behaviour found here in such low $\mu_5$ regime. In particular, we will derive the main phenomenological consequences in terms of observables such as the  energy density, the meson dispersion relation, the quark condensate and the topological susceptibilities. As a first step in this direction, we will concentrate here on the $SU(2)$ effective lagrangian, i.e. only for pion degrees of freedom. The  theoretical tools developed here can be extended to include heavier degrees of freedom, although, as explained above, the main ideas behind considering the $\mu_5$ term are better supported for two light flavours. For that purpose, we will use  effective lagrangian techniques, such as the external source method in the presence of axial and vector sources including the singlet components. That will require the inclusion of new operators and therefore low-energy constants (LEC),  which in particular will allow   to renormalize the different observables. 

The paper is organized as follows. In section \ref{sec:efflag} we will discuss the general formalism used to derive the effective chiral lagrangian in the presence of $\mu_5$. Sections \ref{sec:op2} and \ref{sec:op4} will be devoted, respectively, to the specific $\Od(p^2)$ and $\Od(p^4)$ effective lagrangians, including the new  terms. The analysis of the main phenomenological consequences is carried out in section \ref{sec:pheno}, where we will analyze the pion dispersion relation, the vacuum energy density, chiral symmetry restoration observables, the chiral charge density, the topological susceptibility, the pressure and the speed of sound. We will compare our results with previous works in the literature and we will try to extract as much phenomenological information as possible  from lattice simulations, as well as providing some numerical determinations for the new LEC.

\section{Construction of the Effective Lagrangian}
\label{sec:efflag}

We consider the effective low-energy representation of the generating functional $Z[v,a,s,p,\theta]$ in the case $v=p=0$, $s={\cal M}$ and the axial source given by \eqref{amu5}.  We will also consider $\theta=0$ except for the discussion of the topological susceptibility in section \ref{sec:topsus}.  The construction of the most general, model-independent, effective lagrangian can be carried out within the framework of the external source method, originally introduced in \cite{Gasser:1983yg,Gasser:1984gg} for the $SU(2)$ and $SU(3)$ chiral lagrangian respectively. Within this formalism, the building blocks are the meson fields $U$ and the external sources $v,a,s,p,\theta$, which transform under local transformations of the chiral $SU_L(N_f)\times SU_R(N_f)$ group so that the  action is invariant, up to anomalies. The use of the equations of motion (EOM) to a given order, as well as operator identities, allow to express the lagrangian in terms of the minimum number of operators \cite{Gasser:1983yg,Gasser:1984gg,Scherer:2002tk}. The effective lagrangian formalism should be such that the ultraviolet divergences at a given order can be absorbed by the low-energy constants (LEC) multiplying the different operators, whose finite part can be fixed by the phenomenological analysis of lattice or experimental data. 

This formalism is well defined around the low-energy limit of the theory. Therefore, one has to keep a consistent power counting for derivatives of the meson field and for the external sources in a generic momentum scale $p$. Thus, $d_\mu U, v_\mu, a_\mu =\Od(p)$, $s,p=\Od(p^2)$. Therefore, from \eqref{amu5}, we should keep $\mu_5$ formally as an $\Od(p)$ quantity in the chiral power counting, so that our present treatment is best suited for low and moderate values of $\mu_5$. Although we shall be more precise below about the numerical range of applicability for given observables, we  emphasize that our main purpose is to define a model-independent framework as a benchmark for lattice and theoretical model analyses. 

Two  important additional aspects  should be taken into account in the derivation of the effective lagrangian in the present case: 

First, in the original works \cite{Gasser:1983yg,Gasser:1984gg}, the external sources $v,a,p$ were considered as traceless $SU(N_f)$ fields and therefore those results are not directly applicable to our case in \eqref{amu5}.  The effective lagrangian with those singlet fields included was derived in \cite{Urech:1994hd}  for $SU(3)$ and in \cite{Knecht:1997jw,Meissner:1997fa} for $SU(2)$. The main interest of those works was to apply it to the Electromagnetic (EM) field $v_\mu^0=eA_\mu Q$ with $Q$ the quark  charge matrix, following previous ideas in \cite{Ecker:1988te}.  To construct the lagrangian in that case, the so called "spurion" fields $Q_{L,R}(x)$ are introduced, so that the diagonal part of the QCD lagrangian coupled to external fields is written as

\begin{equation}
{\cal L}_Q=A_\mu\bar q \gamma^\mu\left[Q_L(x)P_L+Q_R(x)P_R\right]q 
\label{diagQlag}
\end{equation}
with $P_{L,R}=(1\mp\gamma_5)/2$ and where $Q_{L,R}(x)$ and $A_\mu$ also transform under chiral transformations, which implies that there will be additional terms in the effective lagrangian depending on  $Q_{L,R}$. The EM case corresponds to $Q_L=Q_R=Q$ and $A_\mu$ the gauge field, although most of the formalism developed in \cite{Urech:1994hd,Knecht:1997jw} is developed for arbitrary $Q_{L,R}$ and $A_\mu$.   Therefore, in our present case, from  \eqref{amu5} we have

 \begin{equation}
 Q_L=-Q_R=\frac{\mu_5}{F}\ID, \quad A_\mu=F\delta_{\mu0}. 
 \label{QLRchoice}
 \end{equation}
 where we have conveniently normalized with $F$, the pion decay constant in the chiral limit, for an easier comparison with previous works dealing with the EM case  \cite{Urech:1994hd,Knecht:1997jw}. Note also that we place the $\mu_5$ term in the $Q$ part, to be consistent with the convention of power counting for the $Q$  fields followed in those works.

 The second observation has to do with covariant derivatives, which for the case \eqref{diagQlag} read

 \begin{eqnarray}
d_\mu U&=&\partial_\mu U-iQ_RA_\mu U+iU Q_L A_\mu
\label{covderu}\\
c_{\mu I} Q_I&=&\partial_\mu Q_I -i [G_{\mu I},Q_I]
\label{covderQ}
\end{eqnarray}
with $I=L,R$ and  $G_{\mu I}=Q_I A_\mu$.  Using the  standard identity  $\tr \left( U^\dagger \partial_\mu U \right)=0$ for $SU(n)$ fields  \cite{Scherer:2002tk},   for the choice \eqref{QLRchoice} we have

\begin{equation}
 \tr \left( U^\dagger d_\mu U \right)=-\tr \left( U d_\mu U^\dagger \right)=2i\delta_{\mu 0} N_f \mu_5
\label{trprop}
\end{equation}

Therefore, the operator  $\tr \left( U^\dagger d_\mu U \right)$ has to be considered as an additional operator for constructing the lagrangian to a given order, unlike in  standard ChPT  or in the EM case $Q_L=Q_R=Q$ where that operator vanishes.

Summarizing, the most general lagrangian at a given order is constructed out of the following fields, where we indicate their chiral power counting:

\begin{eqnarray}
G_{\mu\nu}, \chi, c_{\mu I} Q^I=\Od(p^2), \qquad 
 d_\mu U, Q_{I}=\Od(p), \qquad   U=\Od(1). 
\end{eqnarray}
where $\chi=2B_0(s+ip)$ and $G_{\mu\nu}^I=\partial_\mu G_{\nu I}-\partial_\nu G_{\mu I}-i[G_{\mu I},G_{\nu I}]$, which transform under chiral rotations as   \cite{Urech:1994hd}:

\begin{eqnarray}
U&\rightarrow& g_R U g_L^\dagger\nonumber\\
Q_I&\rightarrow& g_IQ_I g_I^\dagger \quad (I=L,R)\nonumber\\
G_{\mu I}&\rightarrow& g_I G_{\mu I} g_I^\dagger + ig_I \partial_\mu g_I^\dagger \nonumber\\
\chi&\rightarrow& g_R \chi g_L^\dagger \nonumber \\
d_\mu U&\rightarrow& g_R d_\mu U g_L^\dagger\nonumber\\
c_{\mu I} Q_I&\rightarrow& g_I  c_{\mu I}  Q_I g_I^\dagger  \quad (I=L,R) \nonumber\\
G_{\mu\nu}^I&\rightarrow &g_I G_{\mu\nu}^I g_I^\dagger
\label{chiraltrans}
\end{eqnarray}
where $g_{L,R}\in SU(N_f)$ with $N_f=2,3$ light flavors. The lagrangian is constructed demanding the same invariance properties as the QCD one with external sources, namely under chiral rotations, Lorentz covariance and $P,C$ symmetries \cite{Urech:1994hd}:

\begin{eqnarray}
U, \chi &\stackrel{P}{\longleftrightarrow}& U^\dagger, \chi^\dagger, \qquad U, \chi\stackrel{C}{\longleftrightarrow} U^T, \chi^T
\nonumber\\
d_\mu U &\stackrel{P}{\longleftrightarrow}& d^\mu U^\dagger, \qquad d_\mu U \stackrel{C}{\longleftrightarrow}  (d_\mu U)^T \nonumber \\
Q_{L}&\stackrel{P}{\longleftrightarrow}& Q_{R}, \qquad Q_{L}\stackrel{C}{\longleftrightarrow} Q^T_{R}\nonumber\\
G_{\mu}^{L}&\stackrel{P}{\longleftrightarrow}& G^{\mu R}, \qquad G_{\mu}^{L}\stackrel{C}{\longleftrightarrow} \left(G^R_{\mu}\right)^{T}\nonumber\\
c_{\mu L}&\stackrel{P}{\longleftrightarrow}&c_{\mu R}, \qquad c_{\mu L}\stackrel{C}{\longleftrightarrow}c_{\mu R}^T
\label{CPtrans}
\end{eqnarray}

Although in our particular case \eqref{QLRchoice}, we have explicitly broken $P$ and Lorentz covariance in the QCD lagrangian,  keeping $Q_{L,R}$ arbitrary within the external source method, transforming according to \eqref{chiraltrans} and \eqref{CPtrans} under those transformations,  ensures that one is taking into account all possible terms. After using EOM and operator identities,  we will replace in the end the $Q_{L,R}$ fields by  \eqref{QLRchoice}. 

In turn, with this procedure, we will be constructing the most general lagrangian for arbitrary $Q_{L,R}$, which may be useful for other purposes. Such lagrangian will be a generalization of that considered for the EM case in  \cite{Meissner:1997fa,Urech:1994hd,Knecht:1997jw}, which we will reobtain as a consistency check in the case $Q_L=Q_R$. In fact,  some of the needed new LEC multiplying the  lagrangian terms in our case will be  related to the EM LEC in those works. Note that the same procedure can be followed to incorporate other chemical potentials of interest for lattice and heavy-ion phenomenology, such as quark baryon number for $Q_L=Q_R=(\mu_B/F)\ID$, $A_\mu=\delta_{\mu 0} \ID$, charge  for $Q_L=Q_R=(\mu_Q/F)\ Q$, $A_\mu=\delta_{\mu 0} \ID$ or isospin for a combination of the two, or including strangeness $\mu_S$ for three flavors. Those analyses are beyond the scope of this work and will be analyzed elsewhere, being complementary to previous ones in the literature where the low-energy ChPT effective lagrangian framework has been used for analyzing the effect of those chemical potentials   \cite{AlvarezEstrada:1995mh,Son:2000by,Loewe:2002tw,Adhikari:2019mdk, Adhikari:2019mlf}. It is clear that a realistic description of properties relevant to heavy-ion collisions, such as  those commented in section \ref{sec:intro} would require eventually to consider those effects, as well as further observables with respect to the ones studied here. In this respect, although in the present work the only external field we are considering is the axial abelian field in \eqref{QLRchoice} accounting for $\mu_5$, other potentially interesting extension is the inclusion of an external magnetic field through $v_\mu$, which would allow to study the CME in the effective lagrangian context. The modifications on that case would start from the pion propagator itself which is nontrivially modified \cite{Shushpanov:1997sf}.

\section{The leading order $\Od(p^2)$ lagrangian}
\label{sec:op2}

It is not difficult to see that the lowest nontrivial order lagrangian that one can construct for our present case through the previous procedure is the same as in the standard case, i.e., $\Od(p^2)$. Thus, at $\Od(p^0)$,  the only ingredient that we can use is the field $U$, and then all possible terms are constants, independent of $\mu_5$ and then irrelevant for our purposes, while at $\Od(p)$, the only nontrivial operator with the allowed symmetries and arbitrary $Q_{L,R}$ is

\begin{equation}
{\cal L}_1\rightarrow \tr \left(Q_L+Q_R\right)
\end{equation}
which vanishes exactly for the particular choice \eqref{QLRchoice}.

To $\Od(p^2)$, we have, on the one hand, the standard lagrangian in terms of the covariant derivative in \eqref{covderu}, including the additional term coupling the $Q$ and $U$ fields needed to explain the electromagnetic mass difference of pions \cite{Gasser:1983yg,Urech:1994hd,Knecht:1997jw}:

\begin{equation}
{\cal L}_2\rightarrow \frac{F^2}{4}\tr\left[d_\mu U^\dagger d^\mu U  +\chi^\dagger U+\chi U^\dagger\right]+C\tr\left[Q_R U Q_L U^\dagger  \right]
\label{L2lag}
\end{equation}
with $F$ the pion decay constant in the chiral limit and $M^2=2B_0 m$ the tree-level neutral pion mass. We recall that the above equation is valid for arbitrary $Q_L$ and $Q_R$. Taking the EM limit  $Q_L=Q_R=Q$, one can relate the constant $C$ with the EM pion mass difference as  $M_{\pi^+}^2-M_\pi^2=2Ce^2/F^2$ at tree level. 

On the other hand, according to our discussion in section \ref{sec:efflag}, the following operators are also allowed to this order:

\begin{eqnarray}
\tr(U^\dagger d_\mu U)\tr(U^\dagger d^\mu U)&\rightarrow&-4N_f^2\mu_5^2 \nonumber\\
 \tr[Q_L^2+Q_R^2]&\rightarrow& 2N_f \mu_5^2/F^2\nonumber\\
 \tr[Q_L]^2+\tr[Q_R]^2&\rightarrow&2N_f^2 \mu_5^2/F^2\nonumber\\
 \tr[Q_L]\tr[Q_R]&\rightarrow&-N_f^2\mu_5^2/F^2
 \label{extraop2terms}
 \end{eqnarray}
where in the r.h.s. of the above equations we have replaced for those operators our present choice of $Q_{L,R}$  given by \eqref{QLRchoice}, using \eqref{trprop}. Replacing in addition the covariant derivative \eqref{covderQ} in \eqref{L2lag} yields finally for $N_f=2$:

\begin{equation}
{\cal L}_2=\frac{F^2}{4}\tr\left[\partial_\mu U^\dagger \partial^\mu U  +2B_0{\cal M}\left(U+ U^\dagger\right)\right]+2\mu_5^2 F^2\left(1 -Z+\kappa_0\right) 
\label{L2mu5} 
\end{equation}
where the $\kappa_0$ constant accounts for the operators in \eqref{extraop2terms} and we have denoted $Z=C/F^4$ following the notation in \cite{Knecht:1997jw}. Note that numerically $Z\sim 0.8$ \cite{Knecht:1997jw} and therefore we will keep that  contribution in what follows.  

Therefore, at this order, the only modification to the chiral lagrangian is a constant term, which will contribute to the vacuum energy density and to the chiral charge density, as we discuss below. 

Regarding renormalization, it is important to point out that $\kappa_0$ should be finite, since there are no loop divergences to cancel out at this order. We will get back to the renormalization of the new LEC in the following sections.

Finally, we remark that  the equations of motion to $\Od(p^2)$ are $\mu_5$-independent. In our present case they become:

\begin{equation}
(d_\mu d^\mu U^\dagger)U-U^\dagger d_\mu d^\mu U=\chi^\dagger U-U^\dagger \chi+\frac{1}{N_f}\tr\left[U^\dagger\chi-\chi^\dagger U\right]-\frac{4C}{F^2}\left(U^\dagger Q_R U Q_L-Q_L U^\dagger Q_R U\right)
\label{eom} 
\end{equation}
and one can easily check that the all the $\mu_5$ contributions cancel in  \eqref{eom}.

\section{Next to leading order: the $\Od(p^4)$  lagrangian}
\label{sec:op4}

Before discussing the $\Od(p^4)$ we should check first if there are nonvanishing $\Od(p^3)$ terms. The list of all possible terms of that order  allowed by the symmetries is listed in Appendix \ref{sec:op3}. One can readily check that all of those operators vanish for the choice \eqref{QLRchoice}.   

Another important comment regards the Wess-Zumino-Witten (WZW) anomalous part of the lagrangian, which is also $\Od(p^4)$. However,  there are no $\mu_5$-dependent contributions in $SU(2)$, since the WZW lagrangian in that case is independent of the singlet axial field \cite{Kaiser:2000ck}. In the presence of nonzero vector fields, such as the magnetic field needed to study the CME, the WZW would play an essential role \cite{Fukushima:2012fg}.

Let us then follow the same procedure as  before, now to $\Od(p^4)$. The lagrangian to this order will consist of the usual $SU(2)$ terms in \cite{Gasser:1983yg,Gasser:1984gg,Knecht:1997jw} with the covariant derivative $d_\mu$ in \eqref{covderu}, plus new terms constructed out of the $Q$ operators and   the operator $\tr (U^\dagger d_\mu U)$, as commented above. The LEC associated to those new terms will be labelled $k'_i$  and the resulting  lagrangian is given  in eq. \eqref{Lnewsu2} below.  Let us explain the origin of the different terms in that equation. For that purpose, it is convenient to classify the different operators contributing according to the number of $Q$ fields.

 It is not difficult to see that there are no surviving terms with one or three $Q$ fields. These include  $Gdd$ terms like $\tr \left[d_\mu U^\dagger d_\nu U\right] \left[\tr \left(G_{\mu\nu}^L\right)+\tr \left(G_{\mu\nu}^R\right)\right]$, $\tr [U^\dagger d_\mu U] \tr[U^\dagger d^\nu U]  \left[\tr \left(G_{\mu\nu}^L\right)+\tr \left(G_{\mu\nu}^R\right)\right]$ and so on, which in principle could contribute after partial integration moving the derivative acting on the $Q$ fields to the other fields and using the equations of motion \eqref{eom}. However, those terms vanish for our choice of $G_{\mu I}$ after such partial integration.  Terms of the form $\tr \left[G_{\mu\nu}^L d^\mu U^\dagger d_\nu U\right]$, $\tr \left(G^{R,L}_{\mu\nu} G^{\mu\nu R,L}\right)$ and $\tr\left(G^R_{\mu\nu}UG^{L\mu\nu} U^\dagger\right)$ do not contribute either for our present case.

Terms without $Q$ fields include the usual chiral lagrangian at this order \cite{Gasser:1983yg,Gasser:1984gg,Knecht:1997jw} plus new terms which will contain  $\tr (U^\dagger d_\mu U)$.  In Appendix \ref{app:noQ} we list all the possible terms of this type.  Using the equations of motion,  all possible terms with two or more derivatives acting on the same field can be rewritten in terms of those with single derivatives. The latter holds also for terms with derivatives acting on the $\chi$ fields, which after partial integration and the use of the equations of motion can be reduced to those in \eqref{op4termsdermass}. In addition, as discussed in  Appendix \ref{app:op4}, $SU(2)$ operator identities allow to reduce the number of independent terms.

After these considerations, the lagrangian without explicit $Q$ fields is the usual one in  \cite{Knecht:1997jw}, where the $\mu_5$ corrections are those containing the covariant derivative $d_\mu$ in \eqref{covderu}, namely,

\begin{eqnarray}
{\cal L}_4^{0}&=& \frac{l_1}{4}\tr^2\left[ \left(\partial^\mu -2i \mu_5 \delta^{\mu 0}\right)U^\dagger \left(\partial_\mu +2i \mu_5 \delta_{\mu 0}\right)U\right]\nonumber\\
&+&\frac{l_2}{4} \tr\left[ \left(\partial^\mu -2i \mu_5 \delta^{\mu 0}\right)U^\dagger \left(\partial^\nu +2i \mu_5 \delta^{\nu 0}\right)U\right]\tr\left[ \left(\partial_\mu -2i \mu_5 \delta_{\mu 0}\right)U^\dagger \left(\partial_\nu +2i \mu_5 \delta_{\nu 0}\right)U\right]\nonumber\\
&+&\frac{l_3}{16}  \tr^2\left(\chi U^\dagger + U \chi^\dagger \right) +\frac{l_4}{8}\tr\left[ \chi^\dagger U \chi^\dagger U + U^\dagger\chi U^\dagger \chi  \right] \nonumber\\
&+&\frac{l_4}{8} \tr\left[ \left(\partial^\mu -2i \mu_5 \delta^{\mu 0}\right)U^\dagger \left(\partial_\mu +2i \mu_5 \delta_{\mu 0}\right)U\right]\tr\left[(\chi^\dagger U+\chi U^\dagger\right]\nonumber\\
&+& \frac{l_4-l_7}{16} \tr^2\left(\chi U^\dagger - U \chi^\dagger \right) + \frac{h_1+h_3-l_4}{4} \tr\left(\chi^\dagger \chi\right) + \frac{h_1-h_3}{2} \re(\det\chi)
\label{L40Q}
\end{eqnarray}
plus  the new contributions given by the  $k'_{1-5}$ terms in \eqref{Lnewsu2}.  In the $l_4$ term as is customary \cite{Scherer:2002tk,Gasser:1987rb} we have transformed the $l_4$ contribution from  \cite{Knecht:1997jw} by using partial integration, the equations of motion \eqref{eom} and the identity \eqref{trace2dchi1}, which give rise to the terms containing $l_4$ in \eqref{L40Q} plus terms to be absorbed in the $k'_4$, $k'_5$ and $k'_{16}$ operators in \eqref{Lnewsu2}. We do not include in \eqref{L40Q} the $l_5,l_6,h_2$ terms in \cite{Knecht:1997jw}, which contain $G_{\mu\nu}^{L,R}$ and do not contribute here as explained before.

As for terms with two $Q$ fields, the possible contributions are of the form   $ddQQ$ and $\chi QQ$. Once again,  trace identities  can be used to eliminate some of the operators. In particular, we have used eqns.  \eqref{trace2d2Q1}-\eqref{trace2M2Q1} and \eqref{trace2d2Qadd}.  The different operators of this kind contributing to the lagrangian are those multiplied by the LEC $k'_{6-25}$
  in \eqref{Lnewsu2}. We have followed  \cite{Knecht:1997jw} as a guide and for notation, although, as explained, new terms appear with respect to that work which vanish for $Q_L=Q_R$.
  
  Apart from the above, one could in principle have the $(cQ) Qd$ and $(cQ)(cQ)$ operators appearing in Appendix \ref{app:cQQd} and \ref{app:cQcQ} and listed in equations  \eqref{cQQdterms} and 
  \eqref{cQcQterms}  respectively. By partial integration, all those terms can be brought in our present case either to a vanishing contribution or to some of the  $ddQQ$ and $\chi QQ$ contributions already considered.

Finally, we have to consider terms with four explicit $Q$ fields. The relevant trace identities to be used are now \eqref{trace4Q1}-\eqref{trace4Q7} and the operators contributing are those multiplying the new LEC $k'_{26-37}$ in  \eqref{Lnewsu2}.

According to our previous discussion  the $SU(2)$ lagrangian containing the new operators is finally: 
\begin{eqnarray}
{\cal L}_4'&=& k'_{1} \tr(U^\dagger d_\mu U) \tr(U^\dagger d^\mu U) \tr( d^\nu U^\dagger d_\nu U) + k'_{2} \tr(U^\dagger d_\mu U) \tr(U^\dagger d_\nu U) \tr( d^\mu U^\dagger d^\nu U) 
\nonumber\\
&+& k'_{3} \tr(U^\dagger d_\mu U) \tr(U^\dagger d^\mu U) \tr(U^\dagger d_\nu U)  \tr(U^\dagger d^\nu U) + k'_{4} \tr(U^\dagger d_\mu U) \tr(U^\dagger d^\mu U)\tr( \chi^\dagger U + U^\dagger \chi) 
\nonumber\\
&+& k'_{5} \tr(U^\dagger d_\mu U)\tr\left[U^\dagger d_\mu U \left( \chi^\dagger U + U^\dagger \chi \right)\right] 
\nonumber\\
&+&
k'_6F^2\text{tr}\left(d^{\mu}U^{\dagger}d_{\mu}U\right)\left[\text{tr}\left(Q_L^2\right)+\text{tr}\left(Q_R^2\right)\right] 
\nonumber\\
&+&k'_7F^2\text{tr}\left(d^{\mu}U^{\dagger}d_{\mu}U\right)\text{tr}\left(Q_RUQ_LU^{\dagger}\right)  +k'_8F^2\left[\text{tr}\left(d^{\mu}U^{\dagger}Q_RU\right)\text{tr}\left(d_{\mu}U^{\dagger}Q_RU\right)+\text{tr}\left(d^{\mu}UQ_LU^{\dagger}\right)\text{tr}\left(d_{\mu}UQ_LU^{\dagger}\right)\right]
\nonumber\\
&+& k'_9F^2\text{tr}\left(d^{\mu}U^{\dagger}Q_RU\right)\text{tr}\left(d_{\mu}UQ_LU^{\dagger}\right) + k'_{10}F^2\text{tr}\left(d^{\mu}U^{\dagger}d_{\mu}U\right)\left[\text{tr}^2\left(Q_L\right)+\text{tr}^2\left(Q_R\right)\right]
\nonumber\\
&+&  k'_{11}F^2\text{tr}\left(d^{\mu}U^{\dagger}d_{\mu}U\right)\text{tr}\left(Q_L\right)\text{tr}\left(Q_R\right) + k'_{12}F^2\text{tr}\left(\chi^{\dagger}U+U^{\dagger}\chi\right)\left[\text{tr}\left(Q_L^2\right)+\text{tr}\left(Q_R^2\right)\right]
\nonumber\\
&+& k'_{13}F^2\text{tr}\left(\chi^{\dagger}U+U^{\dagger}\chi\right)\text{tr}\left(Q_RUQ_LU^{\dagger}\right) + k'_{14}F^2\text{tr}\left[\left(\chi U^{\dagger}+U\chi^{\dagger}\right)Q_L+\left(\chi^{\dagger} U+U^{\dagger}\chi\right)Q_R\right]\text{tr}\left(Q_L+Q_R\right)
\nonumber\\
&+& k'_{15}F^2\text{tr}\left[\left(\chi^{\dagger} U+U^{\dagger}\chi\right)Q_L-\left(\chi U^{\dagger}+U\chi^{\dagger}\right)Q_R\right]\text{tr}\left(Q_L-Q_R\right)
\nonumber\\
&+& k'_{16}F^2\text{tr}\left[\left(\chi^{\dagger}U-U^{\dagger}\chi\right)Q_LU^{\dagger}Q_RU+\left(\chi U^{\dagger}-U\chi^{\dagger}\right) Q_RUQ_LU^{\dagger}\right]
\nonumber\\
&+& k'_{17}F^2\text{tr}\left(\chi^{\dagger}U+U^{\dagger}\chi\right)\left[\text{tr}^2\left(Q_L\right)+\text{tr}^2\left(Q_R\right)\right]
+ k'_{18}F^2\text{tr}\left(\chi^{\dagger}U+U^{\dagger}\chi\right)\text{tr}\left(Q_L\right)\text{tr}\left(Q_R\right)
\nonumber\\
&+& k'_{19} F^2\tr(U^\dagger d_\mu U) \tr(Ud^\mu U^\dagger Q_R^2 - U^\dagger d_\mu U Q_L^2)
\nonumber\\
&+& k'_{20} F^2\tr(U^\dagger d_\mu U) \tr(U^\dagger d^\mu U) \tr (Q_L^2+Q_R^2)  + k'_{21} F^2 \tr(U^\dagger d_\mu U) \tr(U^\dagger d^\mu U) \tr(Q_R U Q_L U^\dagger)\nonumber\\  
&+& k'_{22} F^2 \tr(U^\dagger d_\mu U) \tr(U^\dagger d^\mu U) \left[  \tr^2 (Q_L) + \tr^2 (Q_R)  \right] + k'_{23} F^2 \tr(U^\dagger d_\mu U) \tr(U^\dagger d^\mu U)  \tr (Q_R) \tr (Q_L) 
\nonumber\\
&+& k'_{24} F^2 \tr(U^\dagger d_\mu U)\left[   \tr(Q_RUd^\mu U^\dagger )\tr(Q_R)-\tr(Q_LU^\dagger d_\mu U)\tr(Q_L) \right]
\nonumber\\
&+& k'_{25} F^2 \tr(U^\dagger d_\mu U)\left[   \tr(Q_RUd^\mu U^\dagger )\tr(Q_L)-\tr(Q_LU^\dagger d_\mu U)\tr(Q_R) \right] 
\nonumber\\
&+& k'_{26} F^4 \left[\text{tr}^2\left(Q_L^2\right)+\text{tr}^2\left(Q_R^2\right)\right] + k'_{27} F^4\text{tr}\left(Q_L^2\right)\text{tr}\left(Q_R^2\right) + k'_{28} F^4\text{tr}\left(Q_L^2+Q_R^2\right)\text{tr}\left(Q_L\right)\text{tr}\left(Q_R\right)
\nonumber\\
&+& k'_{29} F^4\text{tr}\left(Q_L^2+Q_R^2\right)\left(\text{tr}^2Q_L+\text{tr}^2Q_R\right) + k'_{30} F^4\text{tr}\left(Q_R^2-Q_L^2\right)\left(\text{tr}^2Q_L-\text{tr}^2Q_R\right) 
\nonumber\\
 &+& k'_{31} F^4\left[\text{tr}^3\left(Q_L\right)+\text{tr}^3\left(Q_R\right)\right]\text{tr}\left(Q_L+Q_R\right)+k'_{32} F^4\left[\text{tr}^3\left(Q_L\right)-\text{tr}^3\left(Q_R\right)\right]\text{tr}\left(Q_L-Q_R\right)
 \nonumber\\
 &+& k'_{33} F^4\left[\text{tr}^2\left(Q_L\right)\text{tr}\left(Q_R\right)+\text{tr}^2\left(Q_R\right)\text{tr}\left(Q_L\right)\right]\text{tr}\left(Q_L+Q_R\right)
 \nonumber\\
&+& k'_{34} F^4\text{tr}\left(Q_RUQ_LU^{\dagger}\right)\left[\text{tr}^2\left(Q_L\right)+\text{tr}^2\left(Q_R\right)\right] + k'_{35} F^4\text{tr}\left(Q_RUQ_LU^{\dagger}\right)\text{tr}\left(Q_R\right)\text{tr}\left(Q_L\right)
\nonumber\\
&+&
k'_{36} F^4\text{tr}\left(Q_RUQ_LU^{\dagger}\right)\text{tr}\left(Q_R^2+Q_L^2\right) + k'_{37} F^4\text{tr}^2\left(Q_RUQ_LU^{\dagger}\right) 
\label{Lnewsu2}
\end{eqnarray}
so that the full $\Od(p^4)$ lagrangian relevant for our analysis at this order  reads ${\cal L}_4={\cal L}_4^{0}+{\cal L}_4'$ with ${\cal L}_4^{0}$ in \eqref{L40Q}.

The $k'_i$ constants above are dimensionless and can be compared with the EM LEC $k_i$ by taking from the general expressions above $Q_L=Q_R=Q$ with $Q=e \ \diag (2/3,-1/3)$, with $e$ the electric charge,  as considered in \cite{Knecht:1997jw}, which implies in particular $(\tr \ Q)^2=(1/5)\tr(Q^2)$, $\tr(Q)\tr(Q^3)=(9/25)\tr^2(Q^2)$, $\tr(Q^4)=(17/25) \tr^2(Q^2)$. Thus, we get:

\begin{align}
2k'_6+\dfrac{1}{5}\left(2k'_{10}+k'_{11}\right)&=k_1   \hspace{0.5cm}& 2k'_{12}+\dfrac{1}{5}\left(2k'_{17}+k'_{18}\right)&=k_5   \nonumber \\
k'_7&=k_2   \hspace{0.5cm} &  k'_{13}&=k_6  \nonumber \\
k'_8&=k_3   \hspace{0.5cm} &  k'_{14}&=\dfrac{k_7}{2}  \nonumber \\
k'_9&=k_4   \hspace{0.5cm} &  k'_{16}&=k_8 \nonumber \\
2k'_{26}+k'_{27}+\frac{2}{5} k'_{28}+\frac{4}{5}k'_{29}+\frac{4}{25}k'_{31}+\frac{4}{25}k'_{33}&=k_{12}  &\frac{2}{5}k'_{34}+\frac{1}{5}k'_{35}+2k'_{36}&=k_{13} \nonumber\\
k'_{37}&=k_{14}
\label{EMcompsu2}
\end{align}

Note that since the trace condition used in  \cite{Knecht:1997jw} is different from our choice \eqref{QLRchoice},  the combinations of $k'_i$ constants that will appear in the different observables in our case will be different  in general than those in \eqref{EMcompsu2}. Therefore, we will not be able to determine all the LEC appearing in the $\mu_5$-dependent terms in terms of previously known ones. As we will explain below, one can use recent lattice analysis to estimate some of those constants, which one would expect that remain within the same order of magnitude than the $k_i$ ones, from the previous expression \eqref{EMcompsu2}.

Next, let us replace the choice \eqref{QLRchoice}  in the above lagrangian, namely in \eqref{L40Q} and \eqref{Lnewsu2},   in order to get the form of the explicit $\mu_5$-corrections. We obtain

\begin{equation}
\mathcal{L}_4(\mu_5)=\mathcal{L}_4^0 (\mu_5=0)+\kappa_1 \mu_{5}^2\text{tr}\left(\partial^{\mu}U^{\dagger}\partial_{\mu}U\right)
+\kappa_2\mu_5^2\tr\left(\partial_0 U^\dagger \partial^0 U\right)+\kappa_3\mu_{5}^2\text{tr}\left(\chi^{\dagger}U+U^{\dagger}\chi\right)+\kappa_4 \mu_5^4
\label{L4lagmu5}
\end{equation}
with $\mathcal{L}_4^0 (\mu_5)$ in \eqref{L40Q} and

\begin{eqnarray}
\kappa_1&=&4l_1-16k'_1+4k'_6-2k'_7+8k'_{10}-4k'_{11}\label{kappaskprime1}\\
\kappa_2&=&4l_2-16 k'_2\label{kappaskprime2}\\
\kappa_3&=&l_4-16k'_4-8k'_5+4k'_{12}-2k'_{13}+8k'_{15}+8k'_{17}-4k'_{18} \label{kappaskprime3} \\
\kappa_4&=&-128k'_1-128k'_2+256k'_3+32k'_6-16k'_7-32k'_8-16k'_9+64k'_{10}-32k'_{11}+32k'_{19}
-64k'_{20} \nonumber \\
&+&32k'_{21}+32k'_{22}-128k'_{23}+64k'_{24}-64k'_{25}+8k'_{26}+4k'_{27}
-16k'_{28}+32k'_{29}+64k'_{32}-16k'_{34} \nonumber \\
&+&8k'_{35}-8k'_{36}+4k'_{37}
\label{kappaskprime4}
\end{eqnarray}

The method we have followed here to derive the lagrangian \eqref{L4lagmu5}-\eqref{kappaskprime4}  is equivalent to that followed in \cite{Andrianov:2019fwz}, where only the leading LEC in the large-$N_c$ limit are considered, once the proper operator identities for $\mu_5\neq 0$ are taken into account, i.e, those we have used here and collected in Appendix \ref{app:iden}. 

A word about renormalization is also in order here: the $k'_i$ or the $\kappa_i$ LEC have to be renormalized in order to absorb the divergences coming from loops, in the same way as the $l_i$ and the $k_i$ \cite{Gasser:1983yg,Knecht:1997jw}, namely:

\begin{equation}
\kappa_i=\kappa_i^r(\mu)+\beta_i \lambda
\label{kapparen}
\end{equation}
in dimensional regularization (DR), where the superscript ``$r$" denotes the finite part and

\begin{equation}
\lambda=\frac{\mu^{D-4}}{32\pi^2} \left[ \frac{2}{D-4}-\left(\log 4\pi + \Gamma'(1)+1\right)\right]
\label{lambda}
\end{equation}
with  $\mu$ the renormalization scale. 

The values of the $\beta_i$ coefficients will be determined through the analysis of the various observables in the sections below and will imply then conditions on the renormalization of the $k'_i$ showing up in the combinations \eqref{kappaskprime1}-\eqref{kappaskprime4}. In the case of  $\kappa_0$  in \eqref{L2mu5}, since it shows up at $\Od(p^2)$, there is no  counterterm associated to that constant, so that $\beta_0=0$ as emphasized in section \ref{sec:op2}.

\section{Physical consequences}
\label{sec:pheno}

\subsection{Pion dispersion relation}
\label{sec:piondr}

Let us start by analyzing the corrections to the kinetic part of the lagrangian coming from the  $\Od(p^4)$ $\mu_5$-dependent corrections in \eqref{L4lagmu5}. The $\Od(\pi^2)$ part of ${\cal L}_4$ in \eqref{L40Q} and \eqref{L4lagmu5} is given by:

\begin{equation}
\mathcal{L}_4^{\pi^2}=\frac{2l_4 M^2+ 4(\kappa_1+\kappa_2)\mu_5^2}{F^2} \frac{1}{2}\partial_0\pi^a\partial^0 \pi^a - \frac{2l_4 M^2+ 4\kappa_1\mu_5^2}{F^2} \frac{1}{2}\partial_j\pi^a\partial_j \pi^a-\frac{2(l_3+l_4) M^2+ 4\kappa_3\mu_5^2}{F^2} \frac{1}{2}M^2\pi^a\pi_a
\label{L4opi2}
\end{equation}
with $M^2=2B_0 m$.   The  different coefficient for the $\tr (\partial_0 U\partial_0 U^\dagger)$  and $\tr (\partial_i U\partial_i U^\dagger)$ terms translate into different values for the spatial and time components of the pion decay constant, as would be generally expected in a Lorentz covariance breaking  scenario \cite{Pisarski:1996mt}. In our present case we have, up to NLO in ChPT,

\begin{eqnarray}
\left(F_\pi^t\right)^2 (\mu_5)&=&F_\pi^2 (0)  +4(\kappa_1+\kappa_2)\mu_5^2\label{fpit}\\
\left(F_\pi^s\right)^2 (\mu_5)&=&F_\pi^2 (0)  +4\kappa_1 \mu_5^2, \label{fpis} 
\end{eqnarray}
$F_\pi^t$ and $F_\pi^s$ being respectively the generalization of $F_\pi$ for the timelike and spacelike components  of the axial current through the PCAC theorem  \cite{Pisarski:1996mt}. At tree level, they arise directly from  the coefficients of the $\tr (\partial_0 U\partial_0 U^\dagger)$ and $\tr (\partial_i U\partial_i U^\dagger)$ terms in the lagrangian.  In the above equation   we are including  in $F_\pi^2(0)$ the one-loop and $l_i$ standard ChPT corrections \cite{Gasser:1983yg}. Here, it is important to remark that $F_\pi^t/F_\pi^s\neq 1$ implies a reduction of the velocity of propagation of pions \cite{Pisarski:1996mt} as we are about to see. 
 
 When taking into account the above corrections to the derivative terms together with the  $\kappa_3$ correction to the mass term in \eqref{L4lagmu5}, one ends up with the following dispersion relation to this order:
  
\begin{equation} 
p_0^2-\left(1+\delta_s-\delta_t\right) \vert \vec{p} \vert^2 -\left(1+\delta_M-\delta_t\right) M^2 =0 
\label{disprel}
\end{equation} 
where

 \begin{eqnarray}
 \delta_t&=& 2l_4\frac{M^2}{F^2}+A \Delta  +4(\kappa_1+\kappa_2)\frac{\mu_5^2}{F^2}\label{deltat} \\
 \delta_s&=& 2l_4\frac{M^2}{F^2}+A \Delta +4 \kappa_1 \frac{\mu_5^2}{F^2}\ \label{deltas} \\
 \delta_M&=& 2(l_3+l_4)\frac{M^2}{F^2}- B \Delta +4 \kappa_3 \frac{\mu_5^2}{F^2}\ \label{deltaM}
 \end{eqnarray}

Here,  $A,B$ are the coefficients of the loop contributions renormalizing  the $p^2$ and $M^2$ terms of the  dispersion relation at $\mu_5=0$, where $\Delta=G(0)/F^2$ is the tadpole contribution with $G(x)$ the leading-order pion propagator. The divergent part of $\Delta$ is absorbed in DR  in the standard renormalization of the $l_i$ \cite{Gasser:1983yg}.

The two main physical consequences of the above dispersion relation are, on the one hand, a modification of the relativistic pion velocity for massless pions, which at this order is purely a $\mu_5$ effect, namely, 

\begin{equation}
 v_\pi (\mu_5)=\left.\frac{\vert \vec{p} \vert}{p_0}\right\vert_{M=0}=1-\frac{1}{2}(\delta_s-\delta_t)=1+2\kappa_2 \frac{\mu_5^2}{F^2}
 \label{vpi}
 \end{equation}
 
 On the other hand, the loss of Lorentz covariance implies also a different result for the pion mass, depending on whether we take $\vert \vec{p} \vert =0$ (static/pole mass) or $p_0=0$ (screening mass): 
 
\begin{eqnarray}
\left[M_\pi^2 \right]^{pole} (\mu_5)&=& M^2(1+\delta_M-\delta_t)=M_\pi^2 (0) - 4(\kappa_1+\kappa_2-\kappa_3)  \frac{\mu_5^2}{F^2}  M^2
\label{polemass} \\
\left[M_\pi^2 \right]^{scr} (\mu_5)&=&M^2 (1+\delta_M-\delta_s) =M_\pi^2 (0)   - 4(\kappa_1-\kappa_3) \frac{\mu_5^2}{F^2}  M^2
\label{scrmass}
\end{eqnarray}
where $M_\pi^2$ is the $\mu_5=0$ mass at this order including one-loop and $l_i$ terms.
 
 Regarding renormalization, the finiteness of the  observables \eqref{fpit}, \eqref{fpis}, \eqref{vpi}, \eqref{polemass}, \eqref{scrmass} require
 
 \begin{equation}
\beta_1= \beta_2=\beta_3=0
 \label{beta123}
 \end{equation}
 
 The above renormalization conditions highlight the importance of including properly the new terms in the lagrangian \eqref{Lnewsu2}. For instance, the presence of the $k'_2$ term is crucial to guarantee $\beta_2=0$ once the relation with $l_2$ in \eqref{kappaskprime2} is taken into account, since that condition would not be fulfilled by $l_2$ only, whose renormalization is given  in \cite{Gasser:1984gg}. 
 
 As for the numerical values of the LEC $\kappa_1,\kappa_2,\kappa_3$  involved,  there is no  information available from the lattice regarding the pion dispersion relation at $\mu_5\neq 0$. Measuring screening masses for light mesons is, in principle, feasible in the lattice, so it would be useful to have such measurements for $\mu_5\neq 0$ available in the near future.  However, we can have some insight from physical requirements. Hence, requiring that the pion velocity in \eqref{vpi} remains smaller than the speed of light for any $\mu_5$ yields
 
 \begin{equation}
 \kappa_2<0 .\label{kappa2}
 \end{equation}
 
 The additional requirement that the two squared pion masses in \eqref{polemass} and \eqref{scrmass}  remain positive would lead to $\kappa_1-\kappa_3<0$. However, that may be a too restrictive condition, since a decreasing pion mass for low and moderate values of $\mu_5$ does not necessarily imply a tachyonic mode, given that higher order corrections may change this behaviour.  In any case, a tachyonic mode is not necessarily related to an unphysical spectrum \cite{Andrianov:2012dj} and it could indicate for instance  phases of pion condensation  \cite{Son:2000by,Loewe:2002tw}.   We will actually come back to the issue of the sign of that particular LEC combination below, in connection with chiral restoration (section \ref{sec:condensate})  and the chiral charge density (section \ref{sec:chiralcharge}).

\begin{figure}[h]
 \centerline{ \includegraphics[width=0.45\textwidth]{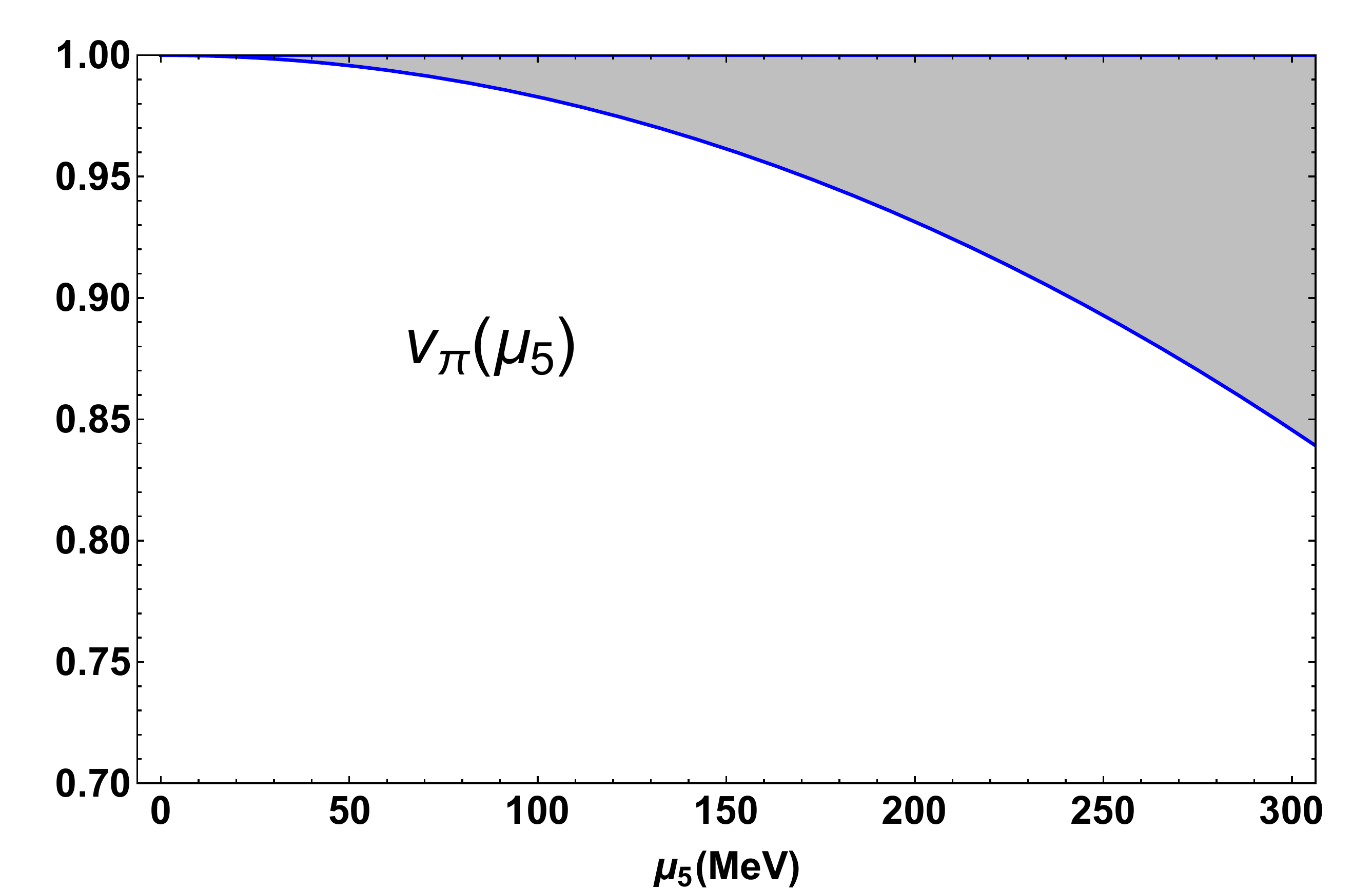}
  \includegraphics[width=0.45\textwidth]{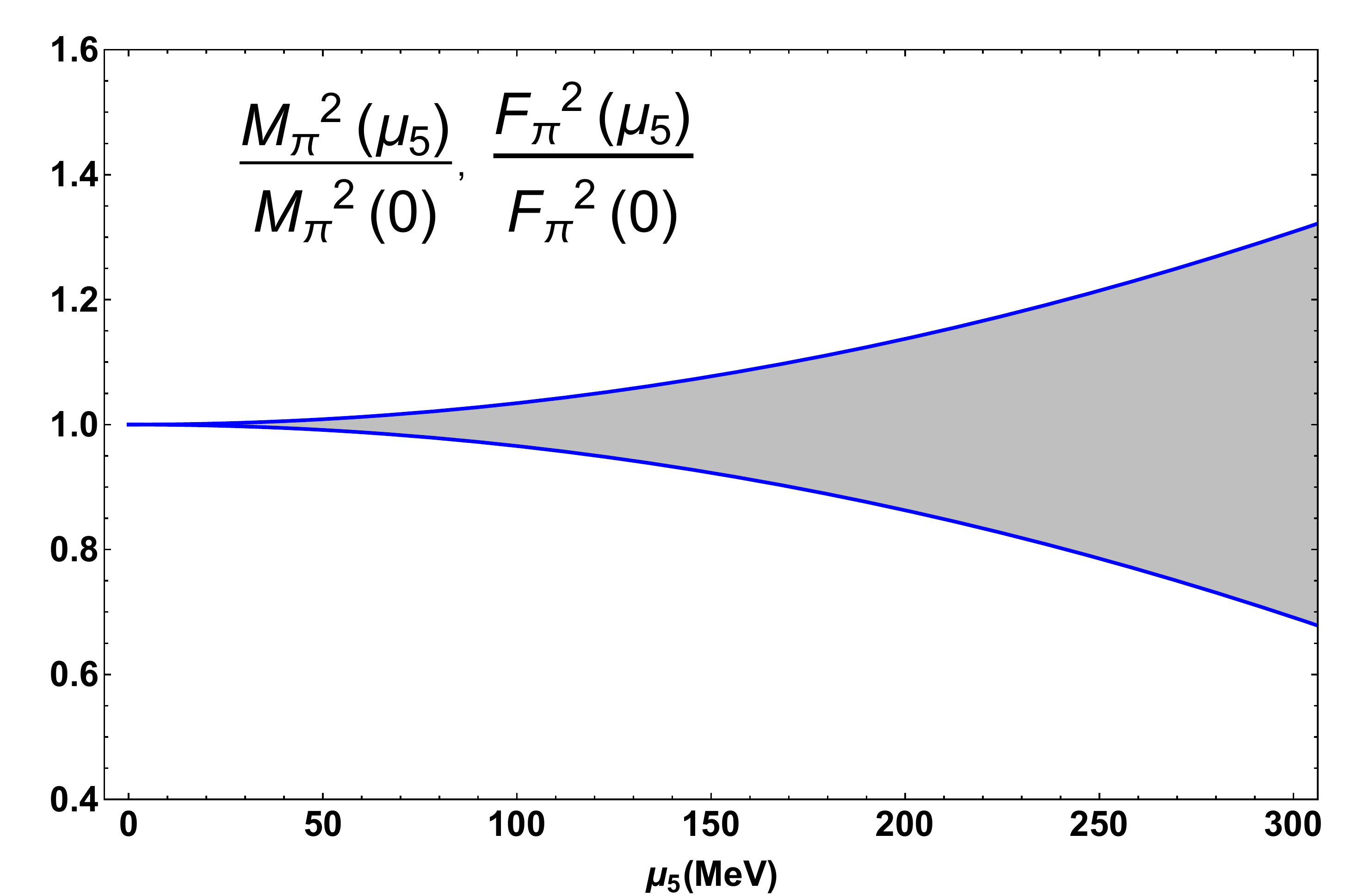}}
  \caption{$\mu_5$ dependence of pion velocity, pion mass and pion decay constant, to leading order in ChPT. The grey bands correspond to the uncertainties of the LEC within natural values, namely $0\leq \kappa_2\leq \frac{1}{16\pi^2}$ in \eqref{vpi}, and $\vert \kappa_1+\kappa_2 \vert \leq \frac{1}{16\pi^2}$, $\vert \kappa_1\vert \leq \frac{1}{16\pi^2}$, $\vert \kappa_1+\kappa_2-\kappa_3 \vert \leq \frac{1}{16\pi^2}$, $\vert \kappa_1-\kappa_3 \vert \leq \frac{1}{16\pi^2}$,   in \eqref{fpit}, \eqref{fpis}, \eqref{polemass} and \eqref{scrmass} respectively.}
  \label{fig:vpimass}
\end{figure}

 Just to obtain a rough estimate of the above results, we plot in Fig. \ref{fig:vpimass} the dependence of $v_\pi$ and $M_\pi^2$ with $\mu_5$ expected within the numerical range of LEC around their so-called natural values $1/(16\pi^2)$ which is their expected size from loop corrections \cite{Urech:1994hd, Knecht:1997jw}.   For the numerical values of the standard low-energy parameters, we will take the recent results quoted in \cite{Aoki:2019cca} and references therein. Thus, we take $F_\pi(0)=92.2$ MeV, its physical value, $F=85.93$ MeV, $M_\pi(0)=140$ MeV, $M=130.96$ MeV.  For the results showed in this figure, we have  replaced for simplicity $F^2\rightarrow F_\pi^2 (0)$ in the right hand side of \eqref{fpit}and \eqref{fpis}, and  $M^2\rightarrow M_\pi^2 (0)$  in \eqref{polemass} and \eqref{scrmass}, which is perturbatively equivalent to this order. Thus, as long as we remain within  natural values for the corresponding LEC involved, the estimated band for $\left[M_\pi^2\right]^{pole,scr}(\mu_5)/M_\pi^2(0)$ and $\left[F_\pi^2\right]^{t,s}(\mu_5)/F_\pi^2(0)$ look the same.

It is worth mentioning also the comparison of our results with previous model analysis. The results in  \cite{Andrianov:2012dj}  within a generalized sigma model, show a decreasing behaviour of the pion mass with $\mu_5$ after diagonalizing a $\pi-a_0$ $\mu_5$-dependent interaction. Actually, tachyonic modes appear in that work for high enough $\mu_5$ and pion momentum.  The results of that paper are compatible with ours for the $\mu_5$ range showed in Fig.\ref{fig:vpimass}. In addition, the increase in $F_\pi^2$ found in \cite{Andrianov:2012dj}  would correspond to positive $\kappa_1$ according to our present analysis and is also numerically compatible with our results in Fig.\ref{fig:vpimass}.  Our analysis for other observables below will show that the lattice results support $\kappa_1>0$, $\kappa_1+\kappa_2>0$, $\kappa_1-\kappa_3>0$, hence compatible with the results in  \cite{Andrianov:2012dj} , although the sign of $\kappa_1+\kappa_2-\kappa_3$ is not determined.

\subsection{The vacuum energy density}
\label{sec:energyden}

Let us analyze the vacuum energy density defined as
\begin{equation}
\epsilon(T,\mu_5)=-(\beta V)^{-1}\log Z (T,\mu_5)
\label{z}
\end{equation}
where $Z(T,\mu_5)=Z(0,a_0=\mu_5\ID,{\cal M},0,0)$ is the Euclidean QCD partition function after the replacement $i\int dx^0\rightarrow \int_0^\beta d\tau$ with $\tau=i x^0$ and $\beta=1/T$ the inverse temperature.  Relevant global observables can be derived from $\epsilon$, such as  the light quark  condensate and the scalar susceptibilty signaling chiral symmetry restoration, as well as the chiral charge density corresponding to the chiral charge \eqref{chiralcharge}. The  $\mu_5$ corrections come from  the lagrangian up to $\Od(p^4)$  given in sections \ref{sec:op2} and \ref{sec:op4} within ChPT. As customary, let us write $\epsilon=\epsilon_2+\epsilon_4 + \cdots$ where $\epsilon_k$ denotes the $\Od(p^k)$ contribution.

\begin{figure}[h]
 \centerline{ \includegraphics[width=0.7\textwidth]{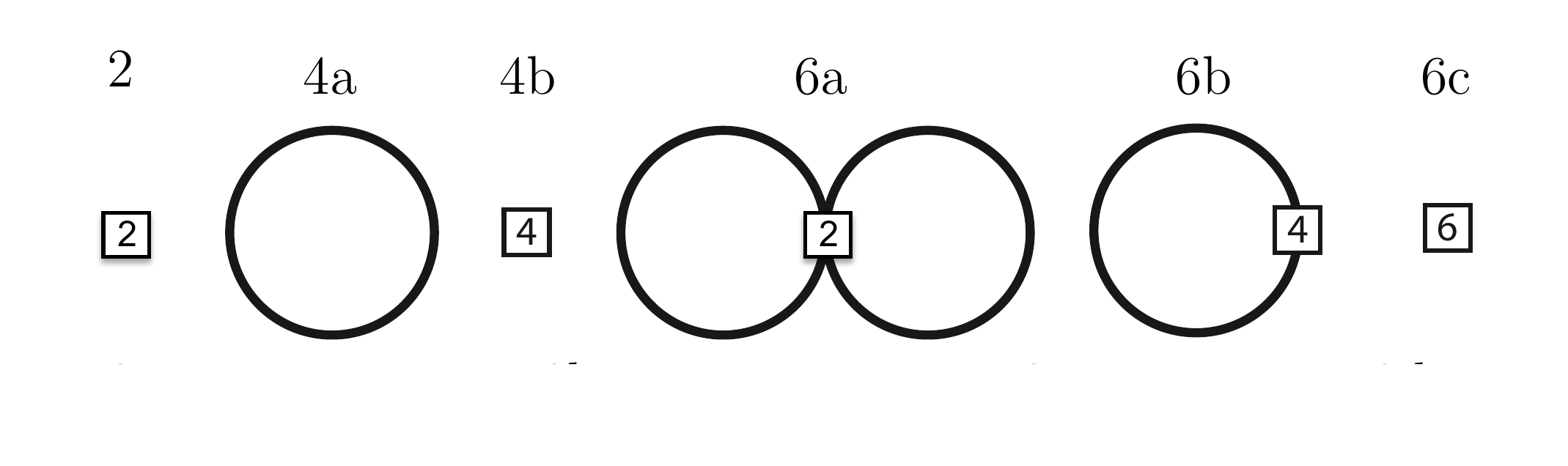}}
  \caption{Diagrams contributing to the energy density up to $\Od(p^6)$}
  \label{fig:diagsenergy}
\end{figure}

At  $\Od(p^2)$, it only contributes the constant (field-independent) part of the ${\cal L}_2$ lagrangian in \eqref{L2mu5}. This is symbolized by the first contribution in Fig.\ref{fig:diagsenergy} labelled ``2" (we follow a similar notation as \cite{Gerber:1988tt}) which yields the following contribution independent of temperature and volume:

\begin{equation}
\epsilon_2 (\mu_5)=-F^2 M^2- 2\mu_5^2 F^2\left(1 -Z+\kappa_0\right) 
\label{e2}
\end{equation}

The $\Od(p^4)$ includes, on the one hand, the contribution from the kinetic $\Od(\pi^2)$ part in \eqref{L2mu5}, which is nothing but the energy density of a free pion gas. This is the closed loop diagram labelled ``4a" in Fig.\ref{fig:diagsenergy}, which is $\mu_5$-independent. On the other hand, the diagram ``4b" refers to the  field-independent terms from ${\cal L}_4$ in \eqref{L40Q} and \eqref{L4lagmu5}. Up to an irrelevant constant, we have in the isospin limit $m_u=m_d$

\begin{eqnarray}
\epsilon_{4a}(T)&=&\frac{3}{2} \frac{T}{V}\sum_n\sum_{k_i} \log\left[\omega_n^2+\omega_k^2\right] 
=3\frac{T}{V}\sum_{k_i} \log\left[1-e^{-\omega_k/T}\right] +M^4\epsilon_{4div}
\underset{V\rightarrow\infty}{\longrightarrow} -\frac{3}{2} g_0(M,T) +M^4\epsilon_{4div}, \label{e4a}\\
\epsilon_{4b} (\mu_5)&=&-(l_3+h_1) M^4-4\kappa_3 \mu_5^2 M^2 - \kappa_4 \mu_5^4 \label{e4b} ,
\end{eqnarray} 
 where we have displayed explicitly the volume and temperature dependence of $\epsilon_{4a}$ as well as its temperature dependence in the infinite volume limit, with $\omega_n=2\pi nT$ ($n\in\IZ$), $\omega_k^2=k^2+M^2$,  $k^2=(2\pi/L)\sum_{i=1}^3 k_i^2$ ($k_i\in\IZ$). The contribution  $\epsilon_{4div}$ in \eqref{e4a} contains a divergent part proportional to $\lambda$ in \eqref{lambda}. In particular, using \eqref{DR1} in Appendix \ref{app:int},

\begin{equation}
\epsilon_{4div}\underset{V\rightarrow\infty}{\longrightarrow} \frac{3}{2}\left[\lambda+ \frac{1}{32\pi^2}\log\frac{M^2}{\mu^2} - \frac{1}{64\pi^2}\right]
\end{equation} 
which is $T$-independent and whose divergent part is also $V$ independent. The divergent $\lambda$ contribution in $\epsilon_{4div}$  cancels  with that of the LEC combination in \eqref{e4b} with the  renormalization of those LEC provided in \cite{Gasser:1984gg}, namely, $l_3=l_3^r(\mu)-\frac{1}{2}\lambda$, $h_1=h_1^r(\mu)+2\lambda$. 

The functions $g_k (M,T)$ above are characteristic of the meson gas. They are defined in \cite{Gerber:1988tt} and satisfy the recurrence relation $g_k=-d g_{k-1}/dM^2$. Specifically, 

\begin{eqnarray}
g_0(M,T)&=& \frac{T^4}{3\pi^2} \int_{M/T}^\infty dx \frac{\left[x^2-(M/T)^2\right]^{3/2}}{e^{x}-1}\label{g0}\\
g_1(M,T)&=&\frac{T^2}{2\pi^2}\int_{M/T}^\infty dx  \frac{\left[x^2-(M/T)^2\right]^{1/2}}{e^{x}-1}
\label{g1}\\
g_2(M,T)&=&\frac{1}{4\pi^2}\int_{M/T}^\infty dx  \frac{\left[x^2-(M/T)^2\right]^{-1/2}}{e^{x}-1}
\end{eqnarray}     
 
The above  three functions are positive, vanish for $T=0$  and increase with $T$ for any mass $M$. In the chiral limit $g_0(0,T)=\pi^2 T^4/45$, $g_1(0,T)=T^2/12$ and $g_2(M\rightarrow 0^+,T)\rightarrow T/(8\pi M)+\Od(\log M^2)$,  $g_3(M\rightarrow 0^+,T)\rightarrow T/(16\pi M)+\Od(\log M^2)$.

Up to $\Od(p^4)$, the only temperature (and volume) dependence of the energy density is contained in the $\mu_5=0$ part, namely in the free pion gas contribution \eqref{e4a}. For some of our subsequent analysis it will be interesting to analyze the nontrivial $T$ and $V$ dependence arising at the  $\Od(p^6)$, although the price will be to introduce more unknown LEC. 

The diagrams contributing to $\epsilon_6$ are also depicted in Fig.\ref{fig:diagsenergy}. The ``6a" contribution stands for the two-loop closed diagram with four-pion vertices coming from ${\cal L}_2$, which is therefore $\mu_5$ independent and has been calculated in  \cite{Gerber:1988tt}:

\begin{equation}
\epsilon_{6a} (T)=\frac{3M^2}{8F^2}\left[G(x=0)\right]^2
\label{e6a}
\end{equation}
with $G$ the euclidean free pion propagator, i.e, 

\begin{equation}
G(x=0)=\frac{T}{V}\sum_n\sum_k \frac{1}{\omega_n^2+\omega_k^2} \underset{V\rightarrow\infty}{\longrightarrow}  G(x=0,T=0)+g_1(M,T)
\end{equation}
where the divergent part is contained in the $T=0$ contribution,  which for $V\rightarrow\infty$ it is given in \eqref{DR0}.

 The contribution from diagram ``6b" stands for  the $\Od(\pi^2)$ part of the ${\cal L}_4$ lagrangian given in \eqref{L4opi2} and is the one containing the combined $\mu_5$ and $T$ dependence. Following similar steps as above, the sum of diagrams 4a and 6b is written in terms of
 
 \begin{eqnarray}
\frac{3}{2}  \frac{T}{V}\sum_n\sum_{k_i} \log\left[a\omega_n^2+b k^2 + c M^2\right]  
\label{e6prev}
\end{eqnarray}

  with
 
 \begin{equation}
 a=1+\frac{2l_4 M^2+ 4(\kappa_1+\kappa_2)\mu_5^2}{F^2} ; \qquad b=1+\frac{2l_4 M^2+ 4\kappa_1\mu_5^2}{F^2}; \qquad c=1+\frac{2(l_3+l_4) M^2+ 4\kappa_3\mu_5^2}{F^2}
 \end{equation}
 
 Up to order $\Od(p^6)$ in the energy density, we can expand the expression in \eqref{e6prev} up to first order in $a-1$, $b-1$, $c-1$, or, equivalently up to first order in $1/F^2$. Thus, performing the Matsubara sums $\sum_n$, we get
 
 \begin{eqnarray}
\epsilon_{6b}(T,\mu_5)&=&\frac{3}{2}  \frac{T}{V}\sum_n\sum_{k_i} \frac{(a-1)\omega_n^2+(b-1)k^2+(c-1)M^2} {\omega_n^2+\omega_k^2}
\nonumber\\
&=& \epsilon_{6bdiv}^{(0)}+ \epsilon_{6bdiv}^{(2)}\mu_5^2+\frac{3}{2} \frac{1}{V}\sum_{k_i} \frac{1}{\omega_k}\frac{1}{e^{\omega_k/T}-1}\left[ (b-a)k^2+(c-a)M^ 2 \right]
\nonumber\\
&\underset{V\rightarrow\infty}{\longrightarrow}&\epsilon_{6bdiv}^{(0)}+ \epsilon_{6bdiv}^{(2)}\mu_5^2+\frac{9}{4} (b-a) g_0(M,T)+ \frac{3}{2}(c-a)M^2g_1(M,T)\nonumber\\
&=&\epsilon_{6bdiv}^{(0)}+ \epsilon_{6bdiv}^{(2)}\mu_5^2-9\kappa_2\frac{\mu_5^2}{F^2} g_0(M,T)+\frac{3\left[l_3M^2+2(\kappa_3-\kappa_1-\kappa_2)\mu_5^2\right]}{F^2}M^2g_1(M,T)
\label{e6b}
 \end{eqnarray}
 where $\epsilon_{6bdiv}^{(0)}$ and  $\epsilon_{6bdiv}^{(2)}$ are $\mu_5$ and $T$ independent divergent contributions in DR. In the $V\rightarrow\infty$ limit they can be obtained from the integral \eqref{DR2} in Appendix \ref{app:int}. We get
 
 \begin{eqnarray}
 \epsilon_{6bdiv}^{(0)} &\underset{V\rightarrow\infty}{\longrightarrow}& 6\frac{M^6}{F^2} l_3 \left[  \lambda + \frac{1}{32\pi^2}\log\frac{M^2}{\mu^2}\right]  
 \label{e6bdiv0} \\
  \epsilon_{6bdiv}^{(2)} &\underset{V\rightarrow\infty}{\longrightarrow}& -3 \frac{M^4}{F^2}\mu_5^2 \left[   \left(4(\kappa_1-\kappa_3)+\kappa_2\right)\left(\lambda + \frac{1}{32\pi^2}\log\frac{M^2}{\mu^2}\right)-\frac{\kappa_2}{64\pi^2} \right]   
 \label{e6bdiv2} 
 \end{eqnarray}
 
 Finally, the ``6c" contribution collects the  field-independent contributions coming from ${\cal L}_6$, which involves three new LEC for the $\mu_5$-dependent part, namely
 
 \begin{equation}
 \epsilon_{6c}(\mu_5)= c \frac{M^6}{F^2}+ \gamma_0 \frac{\mu_5^6}{F^2} + \gamma_1 \frac{\mu_5^4 M^2}{F^2} + \gamma_2 \frac{\mu_5^2 M^4}{F^2}
\label{e6c}
 \end{equation}

 Note that the divergent part coming from the crossed $G(x=0,T=0) g_1$ term in $\epsilon_{6a}$ in eq.\eqref{e6a} cancels with the $l_3$ contribution in \eqref{e6b}, while the $G(x=0,T=0)^2$ term in \eqref{e6a}  and the $\epsilon_{6bdiv}^{(0)}$ term in \eqref{e6b} cancel with the $c$ contribution in \eqref{e6c} with the renormalization $c=c^r(\mu)-6l_3^r(\mu)\lambda+(3/2)\lambda^2$. 
 
 As for the $\mu_5$-dependent contributions, the $\epsilon_{6bdiv}^{(2)}\mu_5^2$ contribution in \eqref{e6b} is absorbed  in the renormalization of  $\gamma_2$  in \eqref{e6c} with 
 
 \begin{equation}
 \gamma_2=\gamma_2^r(\mu) + 3 \left[4(\kappa_1-\kappa_3)+\kappa_2\right]\lambda.
 \label{gamma2ren} 
 \end{equation}
 
 The energy density up to the order calculated here is therefore finite and scale-independent with the renormalization of the LEC that we have just explained, which is a welcomed consistency check. One immediate conclusion is that the LEC $\gamma_0$, $\gamma_1$ in \eqref{e6c} are not renormalized. Also, since the energy density has to remain finite for any $M$, the $\kappa_4$    contribution in  \eqref{e4b}  does not get renormalized either, so that
 
 \begin{equation}
 \beta_4=0
 \end{equation}
 
  In addition, we have checked that the $\mu_5=0$ part of the energy density as given here agrees with \cite{Gerber:1988tt} up to  $\Od(p^6)$. 
 
 In the following sections we will consider different observables that can be obtained from the energy density, and comment on their determination in lattice analysis, which will allow us to gain some information about the $\kappa_i$ LEC involved.

 \subsection{Chiral symmetry restoration: the quark condensate and the scalar susceptibility}
\label{sec:condensate}

The main features of chiral symmetry restoration can be read from the quark condensate and the scalar susceptibility, derived from the vacuum energy density as

\begin{eqnarray}
\condl(T,\mu_5)&=&\langle \bar u u + \bar d d \rangle=\frac{\partial \epsilon(T,\mu_5)}{\partial m}=2B_0 \frac{\partial \epsilon(T,\mu_5)}{\partial M^2} \label{conddef}\\
\chi_S(T,\mu_5)&=& -\frac{\partial \condl (T,\mu_5)}{\partial m}=-2B_0 \frac{\partial \condl (T,\mu_5)}{\partial M^2}
\label{susdef} 
\end{eqnarray}
with $m=m_u=m_d$ in the isospin limit. 

The chiral crossover transition would be characterized by the quark condensate developing a sharp inflection point at the transition temperature $T_c$ and the scalar susceptibility developing a peak at $T_c$, as confirmed by lattice analyses  \cite{Ratti:2018ksb,Bazavov:2018mes}. In the chiral limit, a second-order transition takes place with vanishing condensate and divergent susceptibility \cite{Pisarski:1983ms,Smilga:1995qf}. The ChPT expansion is adequate to provide the low and intermediate temperature behaviour for those quantities, i.e, in the $T$ region where the hadron gas is dominated by the lightest states. More accurate predictions need the inclusion of higher mass states \cite{Ferreres-Sole:2018djq}. However,  the pion gas may still provide a valid qualitative picture of chiral restoration close to the chiral (or infrared) limit  $M\rightarrow 0^+$ which can be formally understood as  $T\gg M$, and where the pions are meant to be the main component responsible for the melting of the quark condensate   \cite{Gerber:1988tt,Smilga:1995qf}. In addition, near the chiral limit, the transition temperature determined in ChPT as the vanishing of the quark condensate coincides with other determinations such as the degeneration temperature of the scalar and pseudoscalar susceptibilities \cite{Nicola:2018vug}.

Our main purpose in this section will  be to study the evolution of chiral restoration for nonzero $\mu_5$, and hence of the quark condensate and the scalar susceptibility and for such purpose the ChPT treatment will be enough to reach the main conclusions, in particular regarding the role of the $\kappa_i$ LEC and the comparison with  lattice analyses.

From the results in the previous section, we see that the first  $\mu_5$ correction to the quark condensate is temperature independent and changes the $T=0$ value of $\condl$ with the term coming from the $\kappa_3$ contribution in \eqref{e4b}, so that the leading and next-to-leading orders for the ChPT quark condensate are given by    \footnote{Throughout this and the following sections we will consider for simplicity the $V\rightarrow\infty$ limit. The finite volume corrections can be introduced along similar lines, replacing the $g_k$ functions by their finite-$V$ counterparts \cite{Gasser:1987ah}},

\begin{eqnarray}
\condl^{LO}&=&-2B_0F^2\label{condlo}\\
\condl^{NLO} (T,\mu_5)&=& -4B_0 M^2\left[l_3^r(\mu)+h_1^r(\mu)-\frac{3}{64\pi^2}\log\frac{M^2}{\mu^2} +2\kappa_3 \frac{\mu_5^2}{M^2} -\frac{3}{4M^2} g_1(M,T) \right]
\label{condnlo}
\end{eqnarray} 

That is, up to this order, the sign of $\kappa_3$ determines whether the quark condensate $\vert \condl \vert$ increases ($\kappa_3>0$) or decreases ($\kappa_3<0$) with $\mu_5$ and so on for the transition temperature $T_c$ estimated from the vanishing of the condensate at this order. As commented in the introduction, lattice analyses at finite temperature favor an increasing behaviour. 
As for the scalar susceptibility, since the $\mu_5$ corrections to the condensate in \eqref{condnlo} are  mass independent, $\chi_S$ is independent of $\mu_5$ at this order. The latter is consistent with the smooth $\mu_5$ dependence observed in the lattice  \cite{Braguta:2015zta}. 

The modification to the $T=0$ condensate value at nonzero $\mu_5$ provided by \eqref{condlo} allows us to make a rough estimate of a typical  $\mu_5$ validity range for the present analysis. Actually, such modification is of the same order as those considered for the pion masses in section  \ref{sec:piondr}, so we expect typically that corrections remain below 20\% for up to $\mu_5\simeq 300-400$ MeV.

Nevertheless, in order to improve the precision in the $T$ and $\mu_5$ range, we consider the next to next to leading order (NNLO) in the energy density, by including the $\epsilon_6$ contributions derived in section \ref{sec:energyden}. The full result for the quark condensate up to that order is given by:

\begin{eqnarray}
\condl^{NNLO} (T,\mu_5)&=&\condl^{NNLO} (T,\mu_5=0)+ \frac{B_0}{F^2}\mu_5^2\left\{   \frac{M^2}{4\pi^2} \left[  3(\kappa_3-\kappa_1) +16\pi^2\gamma_2^r (\mu)\right]   -\frac{3M^2}{8\pi^2} \left[4(\kappa_1-\kappa_3)+\kappa_2 \right]  \log\frac{M^2}{\mu^2}  \right.\nonumber\\
 &-&\left.6\left[   2(\kappa_1-\kappa_3)-\kappa_2 \right]  g_1(M,T)  +12 M^2 \left( \kappa_1-\kappa_3+\kappa_2 \right) g_2(M,T)  +2\gamma_1  \mu_5^2 \right\}
\label{condnnlomu5}
\end{eqnarray}
with
\begin{eqnarray}
\condl^{NNLO} (T,\mu_5=0)&=& \frac{3 B_0 M^4}{1024 \pi ^4 F^2} \left\{128 \pi ^2 \left[16 \pi ^2 c^r(\mu)+l_3^r(\mu)\right]+\left(384 \pi ^2 l_3^r(\mu)+2\right) \log \frac{M^2}{\mu ^2}+3 \log
   ^2\frac{M^2}{\mu ^2}\right\}  \nonumber\\
   &+& \frac{B_0}{F^2} \left\{   \frac{3}{4}\left[g_1(M,T)\right]^2  -\frac{3}{2}M^2 g_1(M,T) g_2(M,T) \right. \nonumber \\
  &+& \left. \frac{3M^2}{32\pi^2}  \left[ 1 + 128 l_3^r(\mu) \pi^2 +2 \log \frac{M^2}{\mu ^2}\right] g_1(M,T) - \frac{3M^4}{32\pi^2} \left[  64  l_3^r(\mu) \pi^2 +  \log \frac{M^2}{\mu ^2}   \right]   g_2(M,T)
  \right\}
  \nonumber\\   
\label{condnnlozero}
\end{eqnarray}

We have checked that the previous expression for $\mu_5=0$ coincides with that given in \cite{GomezNicola:2010tb,GomezNicola:2012uc}, where the  renormalization convention for the LEC is slightly different. 

We see that the leading $T$ and $\mu_5$ dependent corrections to the condensate show up in the $g_1$ and $g_2$ terms in the r.h.s. of \eqref{condnnlomu5},  multiplied by different combinations of the $\kappa_{1,2,3}$ constants, whose sign will determine the condensate evolution with $\mu_5$. Recall that the $\mu_5=0$ condensate is negative, $g_{1,2}$ are positive increasing functions of $T$ and $\kappa_2<0$ according to our discussion in section \ref{sec:piondr}. 

Although, as we have stated above, the ChPT series for the quark condensate cannot provide quantitative reliable predictions around chiral restoration, we expect it to show its main features, especially near the chiral limit, which would correspond to the high temperature $T\gg M$ of the previous expressions.   Thus, in order to provide more quantitative conclusions, let us focus on $T_c(\mu_5)$ determined as the value for which the light quark condensate vanishes.  For that purpose, it is very conveniente  to consider the chiral series for the ratio

\begin{eqnarray}
\frac{\condl (T,\mu_5)}{\condl(0,\mu_5)}&=& 1- \frac{3}{2F^2}g_1(M,T) - \frac{1}{4B_0F^4} \left\{  3g_1(M,T)  \condl^{NLO} (0,\mu_5) \right.
\nonumber\\
&+&\left.2F^2\left[ \condl^{NNLO}(T,\mu_5)- \condl^{NNLO}(0,\mu_5) \right] \right\} + \Od\left(\frac{1}{F^6}\right) \nonumber \\
&=& 1- \frac{3}{2F^2}g_1(M,T) - \frac{3}{4F^4} \left\{ \frac{1}{2}\left[g_1(M,T)\right]^2  -M^2 g_1(M,T) g_2(M,T)\right.\nonumber\\
&-&\left. \left[\frac{M^2}{16\pi^2}\left[ -1+64\pi^2\left(h_1^r(\mu)-l_3^r (\mu) \right)-5 \log \frac{M^2}{\mu ^2}\right]+ 4\mu_5^2 (2\kappa_1-\kappa_2)\right] g_1(M,T) \right.\nonumber\\
&-& \left. \left[ \frac{M^2}{16\pi^2}\left[64\pi^2 l_3^r (\mu) + \log \frac{M^2}{\mu ^2}\right]-8\mu_5^2(\kappa_1 + \kappa_2-\kappa_3)\right] M^2 g_2(M,T) 
\right\}+ \Od\left(\frac{1}{F^6}\right) 
\label{ratiocondmass}
\end{eqnarray}
since the dependence on the $\Od(p^6)$ condensate at $T=0$ cancels, in particular  the constants $c^r$ and $\gamma_2^r$ drop out. Note also that for the above ratio, the $\kappa_i$ dependence reduces to the combinations

\begin{equation}
\kappa_a=2\kappa_1-\kappa_2, \quad \kappa_b=\kappa_1+\kappa_2-\kappa_3,
\label{kappaab}
\end{equation}
where $\kappa_b$  latter is precisely the combination renormalizing the pion pole mass in \eqref{polemass}.  Noter also that the $\mu_5$ dependence is just quadratic, which in particular  implies a quadratic dependence also for $T_c(\mu_5)=T_c(0)\left[1+k\mu_5^2/F^2\right]$ for small $\mu_5$, in accordance with what is found in lattice analysis \cite{Braguta:2015zta,Braguta:2015owi}. 

In the chiral limit, the previous condensate ratio becomes particularly simple, depending only on $\kappa_a$, namely, 

\begin{equation}
\left.\frac{\condl (T,\mu_5)}{\condl(0,\mu_5)}\right\vert_{M=0}= 1 - \frac{T^2}{8F^2}\left[ 1- 2\kappa_a\frac{\mu_5^2}{F^2}\right] -\frac{T^4}{384 F^4} + \Od\left(\frac{1}{F^6}\right)  
\label{ratiocondchiral}
\end{equation}
which yields the following $T_c(\mu_5)$ dependence:

\begin{equation}
\left[T_c(\mu_5)\right]^2 =24F^2\left[ \sqrt{\frac{2}{3}+\left[ 1- 2\kappa_a\frac{\mu_5^2}{F^2}\right]^2}- 1+2\kappa_a\frac{\mu_5^2}{F^2}\right] \qquad (M=0)
\label{tcmu5chiral}
\end{equation}
which for small $\mu_5$ reduces to 

\begin{equation}
T_c(\mu_5)=T_c(0)\left[1+\sqrt{\frac{3}{5}} \kappa_a\frac{\mu_5^2}{F^2}\right] \qquad (M=0)
\label{tcmu5chiralpar}
\end{equation}
with $T_c(0)=2\sqrt{2(\sqrt{15}-3)}F$ in the chiral limit.

\begin{figure}[h]
 \centerline{ \includegraphics[width=0.65\textwidth]{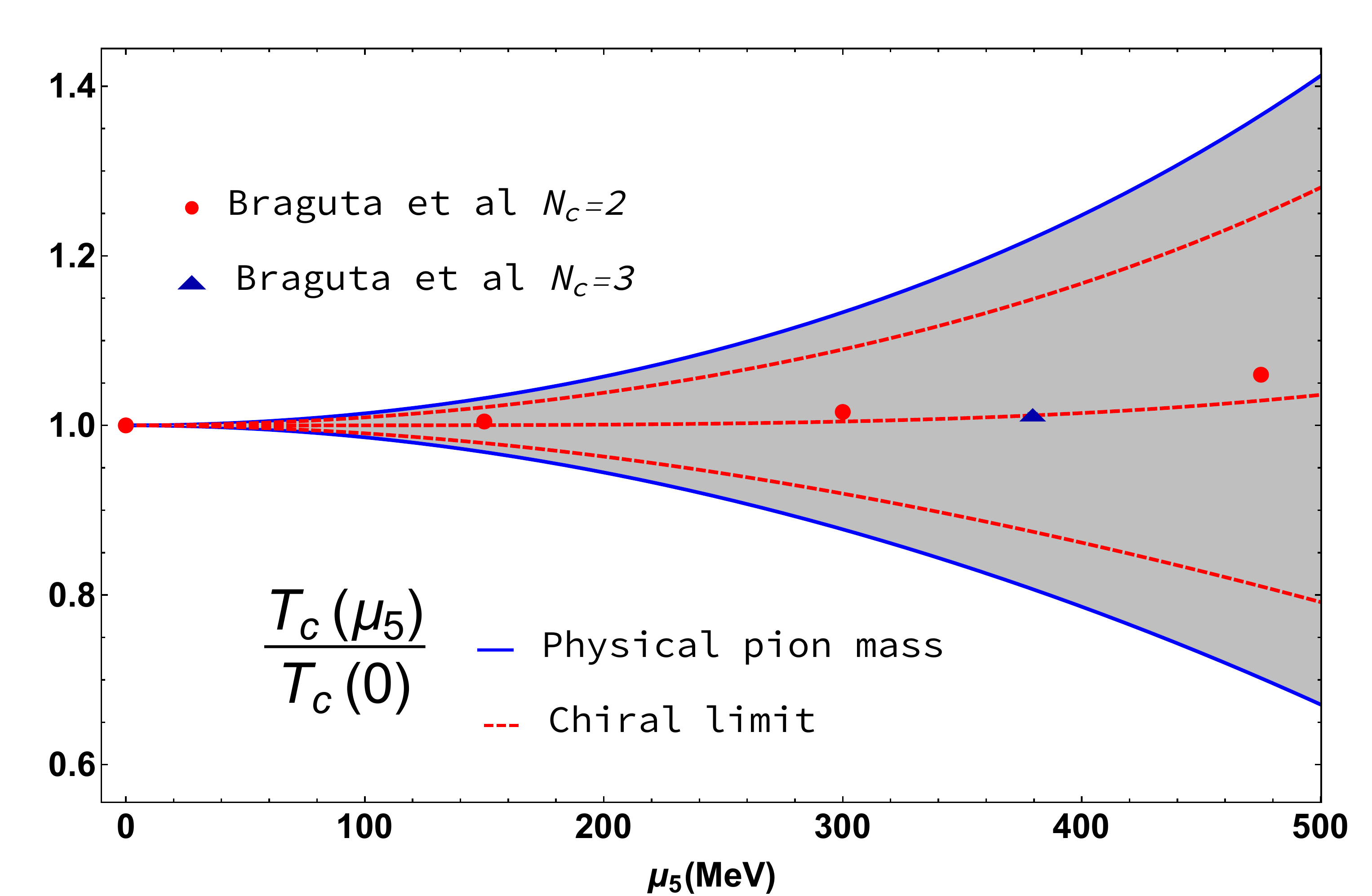}}
  \caption{$\mu_5$ dependence of the chiral transition temperature extracted from the vanishing of the quark condensate. We show the uncertainty bands corresponding to $\vert \kappa_a \vert\leq \frac{1}{16\pi^2}$ and $ \vert \kappa_b \vert \leq \frac{1}{16\pi^2}$ with $\kappa_{a,b}$ defined in \eqref{kappaab} and  where only $\kappa_a$contributes in the chiral limit. We include 
  the lattice points from   \cite{Braguta:2015zta} ($N_c=2$) and  \cite{Braguta:2015owi} ($N_c=3$), as explained in the main text. The corresponding $T_c(0)$ values are 227.1 MeV, and 301.0 MeV for the chiral limit and physical mass curves, and  $T_c(0)=195.8$ MeV for \cite{Braguta:2015zta}.}
  \label{fig:Tc}
\end{figure}

In Fig. \ref{fig:Tc} we plot on the one hand, the expected uncertainty band for $T_c(\mu_5)/T_c(0)$ within the range of natural values for $\kappa_{a,b}$. We take $l_3^r (\mu=770)=  0.21\times 10^{-3}$  \cite{Aoki:2019cca} while for the contact LEC $h_1^r$, which cannot be determined from direct fits, we use the resonance saturation approximation \cite{Ecker:1988te} $H_2^r=2L_8^r$ for those $SU(3)$ LEC, together with the conversion between $SU(2)$ and $SU(3)$ LEC in \cite{Gasser:1984gg} and the $L_6^r$, $L_8^r$ and kaon and eta tree-level masses extracted from  \cite{Aoki:2019cca}. That gives $h_1^r (\mu=770)=6.8\times 10^{-3}$. The same approximation has been used in \cite{GomezNicola:2010tb,GomezNicola:2012uc}.  We show also in that figure the lattice points corresponding to the $N_c=2$ analysis in \cite{Braguta:2015zta}  (with $M_\pi=$ 330 MeV) and the $N_c=3$ one in  \cite{Braguta:2015owi}  (with $M_\pi=$ 550 MeV) which follow similar trends. 

\begin{figure}[h]
 \centerline{ \includegraphics[width=0.45\textwidth]{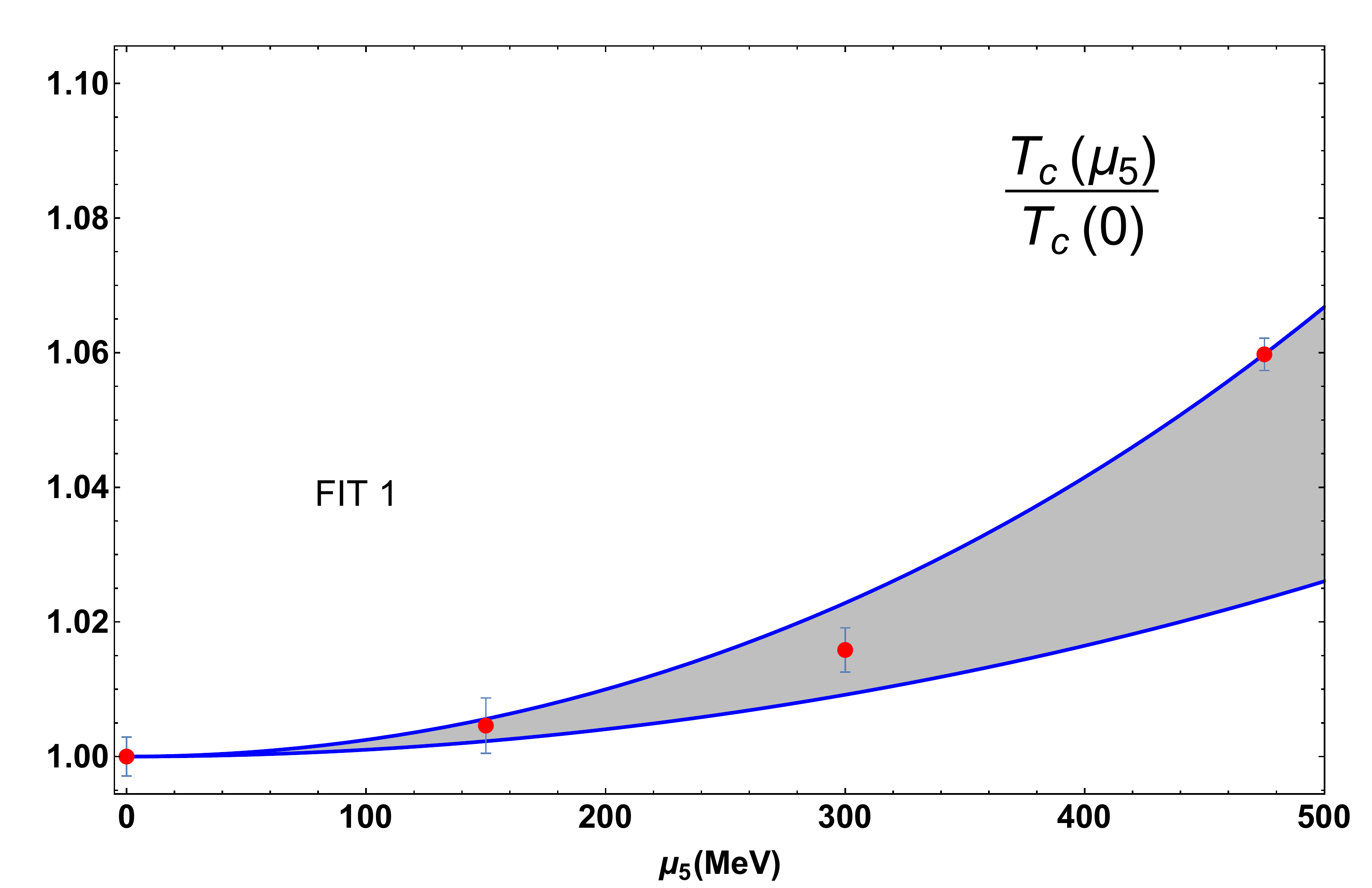}
  \includegraphics[width=0.45\textwidth]{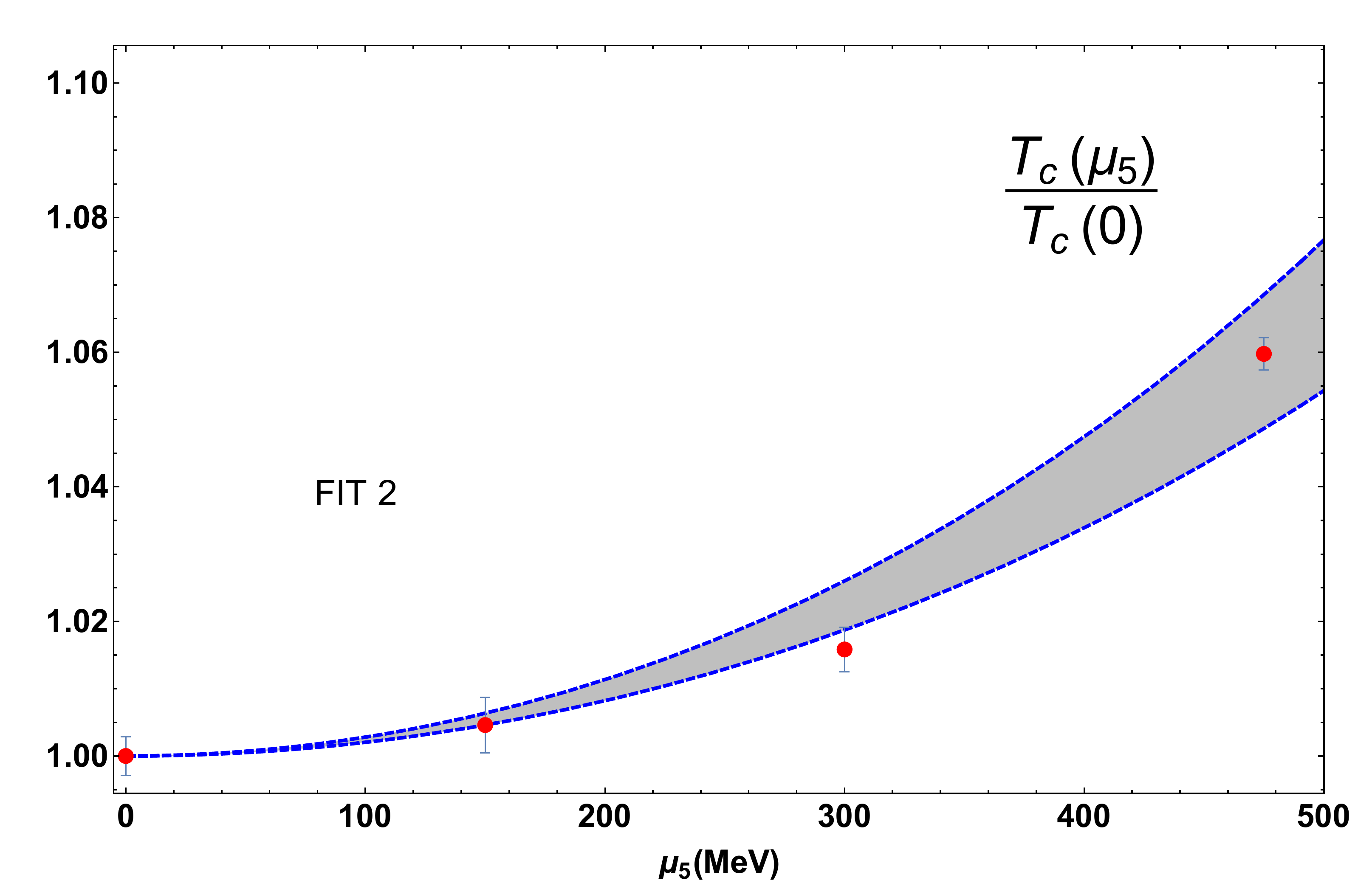}}
   \centerline{ \includegraphics[width=0.45\textwidth]{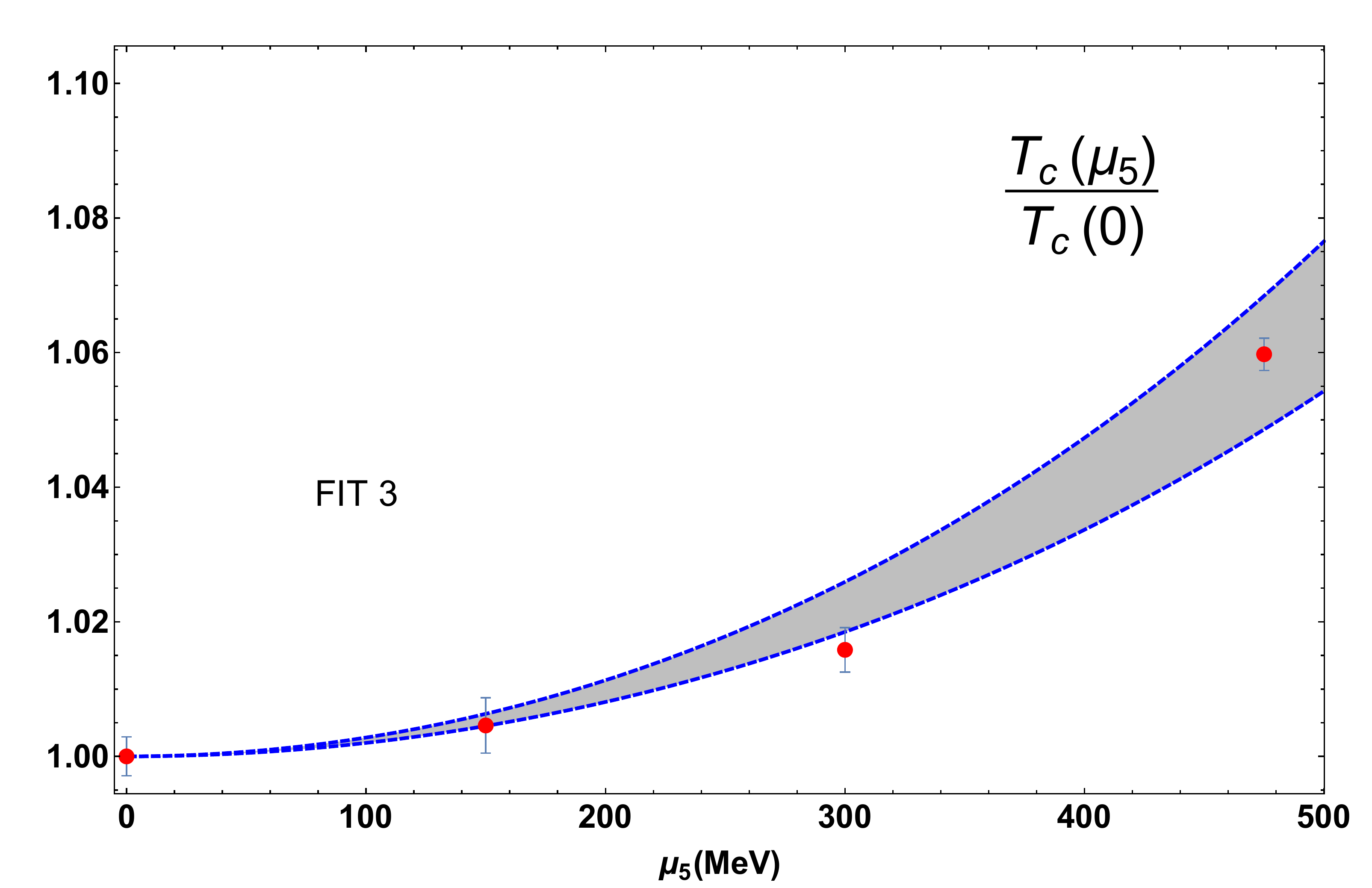}
  \includegraphics[width=0.45\textwidth]{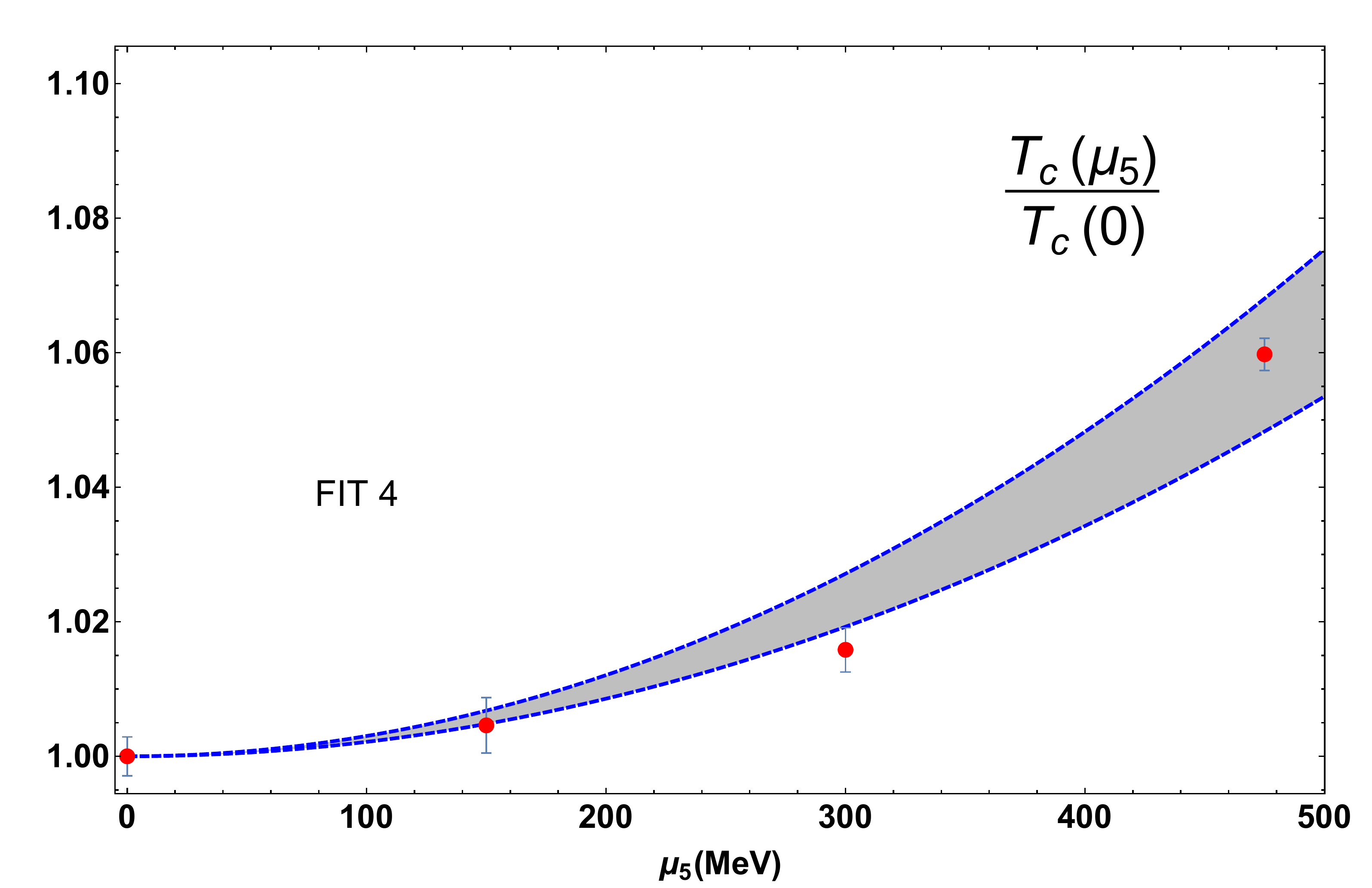}}
  \caption{Fits of $T_c(\mu_5)/T_c(0)$ from the ChPT framework. The lattice points used for the fit are those for  $N_c=2$ in  \cite{Braguta:2015zta} while those for  $N_c=3$ in \cite{Braguta:2015owi} 
  are showed for reference. The  $T_c(0)$ values are the same as in Fig.\ref{fig:Tc}.}
  \label{fig:Tcfits}
\end{figure}

Once we normalize $T_c$ to the corresponding $T_c(0)$ for each case, we see that  ChPT curve for the physical pion mass lies very close to the chiral limit one and that the lattice points clearly fall into the uncertainty given by the natural values range of $\kappa_{a,b}$. This confirms that the main features are captured by the  ChPT approach for the ratio $T_c(\mu_5)/T_c(0)$. Actually, we have performed some fits of the lattice results in order to try to pin down the value of the involved $\kappa_{a,b}$ constants. The results of those fits are given in Fig.\ref{fig:Tcfits}  and in Table \ref{tab:Tcfits}.  The uncertainty bands and parameter errors correspond to the 95\% confidence level of the fits. We have chosen the set of points obtained in  \cite{Braguta:2015zta} since they provide more points in the low $\mu_5$ regime, where our approach is meant to be more applicable. We see that the chiral limit approach in \eqref{tcmu5chiral}, as commented already, yields a very good description of those lattice data for the ratio $T_c(\mu_5)/T_c(0)$.  A fit with only two $\mu_5\neq 0$  points in the chiral limit is shown (fit 1)  while for three points (fit 2) we get a smaller error and still a very good fit. When compared to the massive case (fit 3) fixing the $\kappa_a$ parameter to that of fit 2, we get little sensitivity to $\kappa_b$ (which is compatible with zero)  showing again  that the chiral limit approach with just one parameter $\kappa_a$ is a robust approximation, at least for this particular observable. Actually, setting  the two parameters $\kappa_a,\kappa_b$  free does not improve over the results we present here.  Finally, we also show a fit wit only the parabolic chiral limit $\mu_5^2$ expression in \eqref{tcmu5chiralpar} (fit 4) which  provides a very decent approximation for this $\mu_5$ range.  The values quoted for the $\kappa_a$ parameter in Table \ref{tab:Tcfits} are all compatible within errors and constitute a rather solid prediction of our present analysis, while for $\kappa_b$ the error quoted in fit 3 is narrower than the natural values range but shows a larger uncertainty than $\kappa_a$.

\begin{table}[h]
\begin{tabular}{||l|c|c|c|c|c||}  \hhline{|=|=|=|=|=|=|}
	FIT& $\kappa_a \times 10^3$&$\kappa_b \times 10^3$&$\chi^2/$dof & $R^2$ & \# points $\mu_5\neq 0$\ \\ \hline
	Fit 1 ($M=0$)& 1.7 $\pm$ 0.6& --- &0.01 & 1.00&2 \\ 
	Fit 2 ($M=0$)& 2.3 $\pm$ 0.4& --- & 1.41 & 0.99&3 \\
	Fit 3 & 2.3 (fixed) & 0 $\pm$ 1   &1.36 &0.99&3\\
	Fit 4 ($M=0$ $\Od(\mu_5^2))$&  2.5 $\pm$ 0.4 &  --- &1.85 &0.99&3
	\\ \hhline{|=|=|=|=|=|=|}	\end{tabular}
\caption{Numerical values of the parameters corresponding to the fits in Fig.\ref{fig:Tcfits}. The last column indicates the number of lower $\mu_5$ points considered}
\label{tab:Tcfits}
\end{table}

We end this section by providing the result for the scalar susceptibilty as defined in \eqref{susdef}. The leading nonvanishing order $\Od(1)$ for $\chi_S$ comes from the mass derivative of $\condl^{NLO}$ in \eqref{condnlo} which, as stated, above, is $\mu_5$ independent. The next to leading order in $\chi_S$ is $\Od(1/F^2)$ and corresponds to the derivative of $\condl^{NNLO}$ in \eqref{condnnlomu5}. We get

\begin{eqnarray}
\chi_S(T,\mu_5)&=&\chi_S(T,0)+B_0^2\frac{\mu_5^2}{\pi^2F^2}\left\{\frac{3}{4}\left[6(\kappa_1-\kappa_3)+\kappa_2\right]-8\pi^2\gamma_2^r(\mu)+24\pi^2M^2\left[\kappa_1-\kappa_3+\kappa_2\right]
g_3(M,T)  \right.\nonumber\\
&+&\left. \frac{1}{4}\left[4(\kappa_1-\kappa_3)+\kappa_2\right]\left[-48\pi^2g_2(M,T)+3\log \frac{M^2}{\mu ^2}      \right]  \right\} +\Od\left(\frac{1}{F^4}\right)
\label{chis}
\end{eqnarray}
with $\chi_S(T,0)$ given in  \cite{GomezNicola:2010tb,GomezNicola:2012uc}. 
As explained above, it is not obvious how to extract useful information about the transition from $\chi_S$ calculated in ChPT , since the ChPT approach solely does not reproduce the expected maximum around $T_c$. However, retaining only the leading terms in the chiral or infrared limit $M\rightarrow 0^+$, the divergent contribution for $\chi_S$ is meant to carry the essential information regarding chiral restoration  \cite{Smilga:1995qf}. In that limit, we obtain

\begin{equation}
\chi_S(T,\mu_5)=\frac{3B_0^2}{4} \frac{T}{\pi M}\left\{ 1+  \frac{1}{F^2} \left[  \frac{T^2}{16}-6\left(\kappa_1-\kappa_3\right)\mu_5^2  \right]    \right\}  +\Od(\log M^2) +\Od\left(\frac{1}{F^4}\right)
\label{chischiral}
\end{equation}

Interestingly, the coefficient that regulates the dependence of $\chi_S$  with $\mu_5$ near the chiral limit is precisely $\kappa_1-\kappa_3$, appearing in the pole and screening pion mass corrections in \eqref{polemass} and \eqref{scrmass}. The lattice data indicate that $\chi_S$ decreases with $\mu_5$ below the transition \cite{Braguta:2015zta}, consistently with the peak of $\chi_S$ signaling the transition moving towards higher values of $T$. This favors the sign  $\kappa_1-\kappa_3 >0$, in agreement with the results in \cite{Andrianov:2012dj} on the pion mass. Once more, it becomes clear that lattice measurements on the pion masses would be important to determine at least the sign of the $\mu_5$ corrections, connected as we have just seen with the chiral restoring behaviour. 
The temperatures considered in  \cite{Braguta:2015zta} are too high to trust a fit based on our previous ChPT chiral susceptibility for a pion gas, as commented above. 

\subsection{The chiral charge density and $\mu_5=0$ stability}
\label{sec:chiralcharge}

Another quantity whose $\mu_5$ dependence has been studied in the lattice is the chiral charge density \cite{Yamamoto:2011ks,Astrakhantsev:2019wnp}, defined as

\begin{eqnarray}
\rho_5 (T,\mu_5)&=& \langle  J_5^0 \rangle = \langle \bar q \gamma^0 \gamma_5 q \rangle =- \frac{\partial \epsilon(T,\mu_5)}{\partial \mu_5}   \label{rho5def}
\end{eqnarray}

From our expressions in section \ref{sec:energyden} we  get, up to $\Od(p^6)$ in the free energy,

\begin{equation}
\rho_5 (T,\mu_5)=\rho_5^{(1)}(T) \mu_5 + \rho_5^{(3)} \mu_5^3  + \rho_5^{(5)} \mu_5^5 + \Od(1/F^4),
\label{rho5chpt}
\end{equation}
with
\begin{eqnarray}
\rho_5^{(1)} (T) &=& 4F^2 \left(1-Z+\kappa_0+2\kappa_3 \frac{M^2}{F^2}\right) 
+\frac{1}{32\pi^2 F^2}\left\{  6\left[4(\kappa_1-\kappa_3)+\kappa_2\right]M^4\log \frac{M^2}{\mu ^2}   + 576\pi^2 \kappa_2g_0(M,T)   \right.\nonumber\\
&+&\left.
 384 \pi^2 \left(\kappa_1-\kappa_3+\kappa_2\right) M^2 g_1(M,T)  - M^4\left[3\kappa_2+64\pi^2\gamma_2^r(\mu) \right] \right\},
   \label{rho51} \\
\rho_5^{(3)}&=& 4\left(\kappa_4 -\frac{M^2}{F^2}\gamma_1 \right), \label{rho53} \\
\rho_5^{(5)}&=& -6\frac{\gamma_0}{F^2}\label{rho55} 
\end{eqnarray}

The above expressions provide the ChPT prediction for the chiral charge density, which as in the previous observables discussed, should be applicable at low and moderate values of $T$ and $\mu_5$. Note that the thermal functions above are multiplied by precisely the same $\kappa_i$ combinations showing up in the pion dispersion relation analysis in section \ref{sec:piondr}  and that, according to our analysis in that section and in section \ref{sec:condensate}, $\kappa_2<0$, and therefore the thermal contribution  increases the coefficient of the linear term above in the chiral limit. The sign of $\kappa_b=\kappa_1+\kappa_2-\kappa_3$ is not clear, as we have discussed in section \ref{sec:condensate}.  Note that the chiral charge density satisfies $\rho_5(\mu_5=0)=0$, which is consistent with the Vafa-Witten theorem stating that parity cannot be spontaneously broken  in the absence of axial sources \cite{Vafa:1984xg}. A way out of this theorem is to consider for instance nonzero baryon density  \cite{Andrianov:2013dta,Andrianov:2009pm}.

From the recent lattice analysis in \cite{Astrakhantsev:2019wnp} performed for $N_f=2$, one concludes  that for low and moderate values of $\mu_5$, $\rho_5(\mu_5)$ is very insensitive to the quark masses and its behaviour is pretty much dominated by the linear term $\rho_5\propto \mu_5$. This is consistent with our results  above since the dominant $\Od(F^2)$ term shows up only in the linear term \eqref{rho51} and is mass independent. Actually, in \cite{Astrakhantsev:2019wnp}, that term is estimated simply as $4f_\pi^2$, which is adequate regarding its order of magnitude, but as we have seen in detail here, at that order the contributions from the $Z$ and $\kappa_0$ terms have to be considered in addition to that of the pion decay constant.  Another relevant comment in this context is that the temperature and volume corrections might have to be considered in future lattice analyses. Actually, in  \cite{Astrakhantsev:2019wnp} the total four-volume of the lattice is fixed at around $(1.7 \ \mbox{fm})^4$ which would amount to an effective temperature $T\sim 116$ MeV.  According to our expressions above, that would affect mostly the linear term $\rho_5^{(1)}$, although we expect that the corrections are small at those temperatures since they appear at the NNLO $\Od(1/F^2)$ (see below). The lattice analysis in \cite{Yamamoto:2011ks} is performed at very  high temperatures $T> 400$ MeV and then our present ChPT-based analysis is less appropriate to describe those results. Nevertheless, the almost linear growth of $\rho_5(\mu_5)$ also holds in \cite{Yamamoto:2011ks}.

Let us then perform a fit of $\rho_5(\mu_5)$ to the lowest values of $\mu_5$ provided in \cite{Astrakhantsev:2019wnp}, similarly to what we did in section \ref{sec:condensate} for the critical temperature ratio. For that purpose, and in view of our previous discussion, it makes sense to consider only the chiral limit of our previous expressions for $\rho_5$, since we do not expect to extract any useful information about the constants multiplying the mass terms, to which $\rho_5$ is much less sensitive.  We get

\begin{equation}
\left.\rho_5 (T,\mu_5)\right\vert_{M=0}=4F^2 \mu_5\left(1-Z+\kappa_0  + \kappa_2\frac{\pi^2T^4}{10F^4} \right) +4\kappa_4 \mu_5^3-6\frac{\gamma_0}{F^2}\mu_5^5 + \Od(1/F^4).
\label{rho5chlim}
\end{equation}

\begin{figure}[h]
 \centerline{ \includegraphics[width=0.45\textwidth]{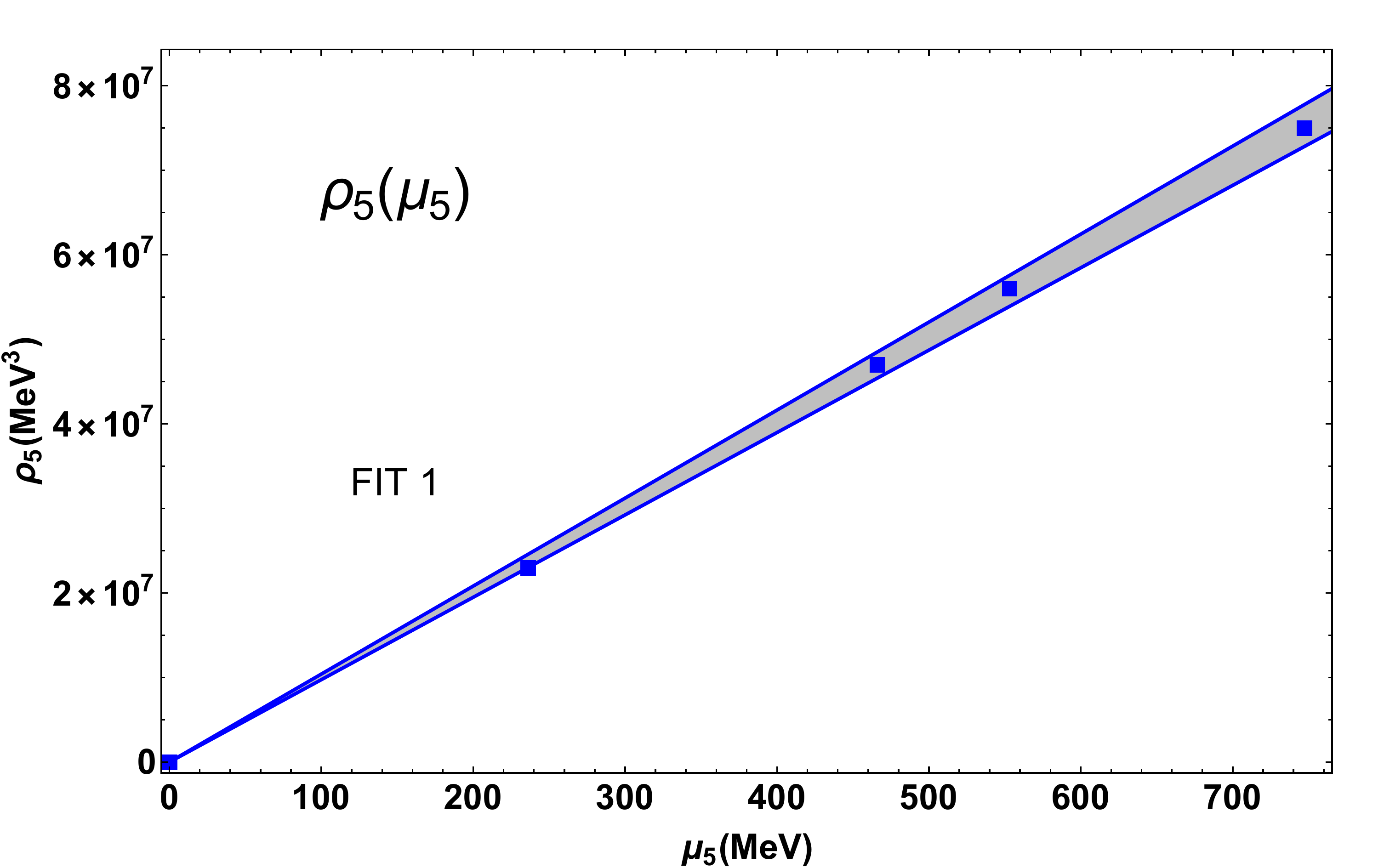}
  \includegraphics[width=0.45\textwidth]{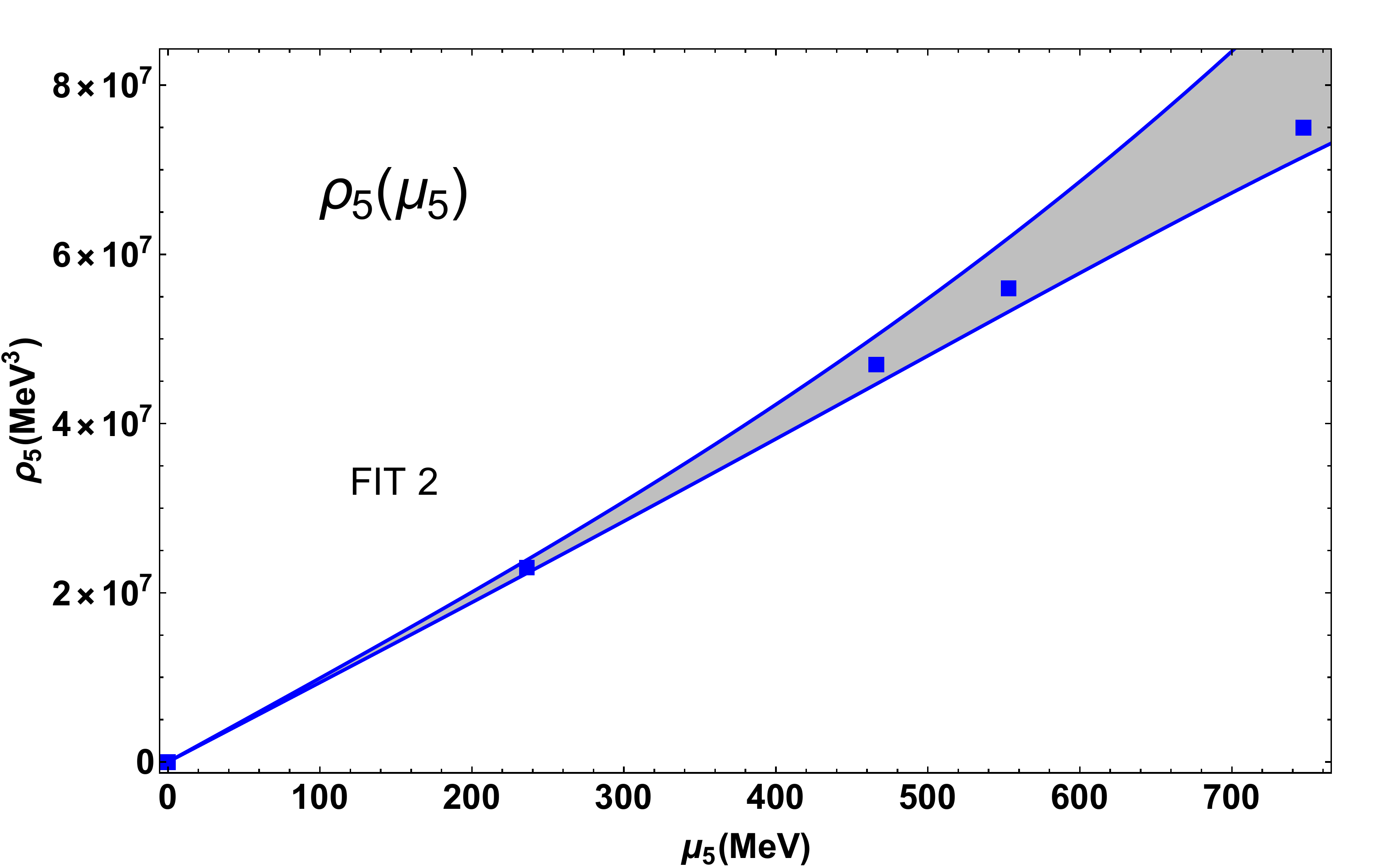}}
  \caption{Fits of $\rho_5(\mu_5)$ using the chiral limit expression \eqref{rho5chlim} for $\kappa_2=0$.   The lattice points used for the fit are those in \cite{Astrakhantsev:2019wnp}.}
  \label{fig:rho5fits}
\end{figure}

With the previous expression, we have performed the fits showed in Fig.\ref{fig:rho5fits} and Table \ref{tab:rho5fits}, for which the uncertainty bands and parameter errors correspond to the 95\% confidence level and we have used the estimate $Z\sim 0.8$ in \cite{Knecht:1997jw}. Lattice errors are not provided in \cite{Astrakhantsev:2019wnp} for $\rho_5$. In both fits, we have not included the temperature dependent term proportional to $\kappa_2$ in \eqref{rho5chlim}. Actually, we get $\kappa_2\frac{\pi^2T^4}{10F^4} \sim 6\times 10^{-3} (1-Z+\kappa_0)$ with the $\kappa_0$ values quoted in Table \ref{tab:rho5fits} and setting the natural value  $\kappa_2=1/(16\pi^2)$ and $T=116$ MeV (as corresponds to \cite{Astrakhantsev:2019wnp}). Thus, for these fits it is completely justified to ignore the volume or temperature dependence, as expected. That contribution is relatively much smaller than the typical error quoted in Table \ref{tab:rho5fits} and therefore no useful conclusion about that value of $\kappa_2$ can be inferred from this analysis.

\begin{table}[h]
\begin{tabular}{||l|c|c|c|c|c||}  \hhline{|=|=|=|=|=|=|}
	FIT& $\kappa_0 $&$\kappa_4 \times 10^3$&$\gamma_0\times 10^5$ & $R^2$ & \# points $\mu_5\neq 0$\ \\ \hline
	Fit 1 & 3.2 $\pm$ 0.1& 0 (fixed) & 0 (fixed) & 0.99&3 \\ 
	Fit 2 & 3.1 $\pm$ 0.1& 7.1 $\pm$ 3.6   &4.6 $\pm$ 2.4 &0.99&4
	\\ \hhline{|=|=|=|=|=|=|}	\end{tabular}
\caption{Numerical values of the parameters corresponding to the fits in Fig.\ref{fig:rho5fits}. The last column indicates the number of lower $\mu_5$ points considered}
\label{tab:rho5fits}
\end{table}

The results show that the simple linear dependence setting $\kappa_4=\gamma_0=0$ (fit 1) already fits very well the lowest $\mu_5$ lattice points. The prediction for $\kappa_0$ is consistent with the fit allowing the three parameters $\kappa_0,\kappa_4,\gamma_0$ to be free (fit 2) which allow to include an additional point, expected to be more sensitive to the nonlinear dependence.  Recall that the numerical value for $\kappa_0$ is not expected to lie within natural values $1/(16\pi^2)$ since it is a low-energy constant of the ${\cal L}_2$ lagrangian. As a consequence of the dominance of the linear term, for the  numerical values of $\kappa_2$ and $\gamma_0$ in fit 2 are affected by larger errors. Our ChPT-based approach essentially captures then the main features of the lattice results.

A relevant issue related to the previous $\rho_5$ analysis has to do with the stability of the $\mu_5=0$ solution. One may wonder whether the free energy admits minima with $\mu_5\neq 0$, which would define a characteristic ''expected" value for the axial chemical potential, at least within the present low-energy approach.   The  temperature and volume dependence of such solution  is relevant since it could be achieved within a heavy-ion collision and/or lattice environment.

Thus, writing the free energy as  $\epsilon(\mu_5)=\epsilon (0)  -\frac{1}{2}\rho_5^{(1)} \mu_5^2 - \frac{1}{4}\rho_5^{(3)} \mu_5^4 + \cdots$ with the $\rho_5^{(i)}$  coefficients defined in our previous analysis of the chiral charge density, the behaviour of $\epsilon(\mu_5)$ around $\mu_5=0$ is controlled by the coefficient of the $\mu_5^2$ term, which according to our previous discussion satisfies $\rho_5^{(1)} >0$ to ensure the growing behaviour of $\rho_5(\mu_5)$ observed in the lattice, which appears to be pretty independent of the temperature and volume.

Therefore, $\epsilon (\mu_5)$ would have a maximum at $\mu_5=0$, which opens up the interesting possibility that the energy density develops a minimum at a nonzero (and not very large) value of $\mu_5$, which would require $\rho_5^{(3)}<0$ and would be given by $\left[\mu_5^2\right]_{min}=-\rho_5^{(1)} /\rho_5^{(3)}$.  At the order we are calculating here,  $\rho_5^{(3)}$ is dominated by  $\kappa_4$ in \eqref{rho53} since the $\gamma_1$ contribution is expected to be suppressed, as well as the $\gamma_0$ $\Od(\mu_5^6)$ term in the free energy given by \eqref{rho55}. On the other hand, our previous fits to lattice data suggest $\kappa_4>0$. In any case, even setting a negative value for $\kappa_4$ of the expected size $10^{-3}$,  would give $\left[\mu_5\right]_{min}\simeq 30F \simeq 2600$ MeV, much higher than the typical applicability range of our present approach.  Therefore, our present analysis does not favor a $\mu_5\neq 0$ minimum for the free energy for low and moderate values of $\mu_5$. This conclusion is independent of temperature and volume, at least at the order considered here.

\subsection{The topological susceptibility}
\label{sec:topsus}

The topological susceptibility for $\mu_5\neq 0$ has also been recently analyzed in the lattice  \cite{Astrakhantsev:2019wnp} and  it might provide additional information about the $\kappa_i$ LEC. it is defined as

\begin{equation}
\chi_{top}=\left.\frac{\partial^2\epsilon(\theta)}{\partial^2\theta}\right\vert_{\theta=0}
\label{topsusdef}
\end{equation}
where $\epsilon(\theta)$ is the vacuum energy defined through \eqref{z} when the $\theta$-term is included in the QCD lagrangian.   From our $\mu_5$-dependent effective lagrangian, we can  calculate the topological susceptibility by noting that a constant $\theta$ term amounts to a complex quark mass matrix according to \eqref{zqcdtransf1}.  This is actually the way that $\chi_{top}$ has been  calculated within the effective theory ChPT framework for $SU(2)$ and $SU(3)$ \cite{Mao:2009sy,Guo:2015oxa} . A systematic study within $U(3)$ ChPT including the $\eta'$ has been recently carried out in \cite{Nicola:2019ohb}. It is important to remark that the quark mass dependence of $\chi_{top}$ favors the dominance of light quarks, being proportional to $M_\pi^2$ in the  $SU(2)$ limit $m_{u,d}\ll m_s$.  Actually, this would explain why the ChPT predictions remain close to the lattice analysis.

Following then the procedure in \cite{Mao:2009sy,Guo:2015oxa}, one considers a nontrivial unitary vacuum configuration different from $U=1$ where $U$ is the Goldstone boson matrix field, namely 
  $U_0=\mbox{diag} \left(e^{i\varphi},e^{-i\varphi}\right)$ where $\varphi$ minimizes $\epsilon(\theta)$ with $m_u\neq m_d$. Now,  the only $\mu_5$-dependent term up to ${\cal L}_4$ proportional to the quark mass is the $\kappa_3$ contribution in \eqref{L4lagmu5}, which has precisely the same form as the mass term in ${\cal L}_2$ in \eqref{L2lag}. Therefore, at this order this amounts to a redefinition $F^2\rightarrow F^2+4\mu_5^2 \kappa_3$ in  $\epsilon_2(\theta)=-F^2M^2\left[ 1+ \frac{1}{8}(\hat\epsilon^2-1)\theta^2 + \Od(\theta^4) \right]$  with $\hat\epsilon=(m_u-m_d)/(m_u+m_d)$  \cite{Guo:2015oxa} so that we get
  
  \begin{equation}
  \chi_{top}(\mu_5)= \chi_{top}(0)+\kappa_3 \mu_5^2 M^2 (1-\hat\epsilon^2) + \Od(p^6/F^2)
  \end{equation}
  with $ \chi_{top}(0)$ given in \cite{Mao:2009sy,Guo:2015oxa}. 
  
  Note that the above results imply in particular that for $\theta=0$ the solution $\varphi=0$ is still  the vacuum energy minimum at $\mu_5\neq 0$, which is compatible with the absence of a pion condensate at the order we are calculating here.

  Therefore, the dependence of $\chi_{top}$ with low and moderate $\mu_5$ is controlled by the $\kappa_3$ constant. We could try to fix it with lattice data as in previous sections. However, the only available results in \cite{Astrakhantsev:2019wnp}  lack of a solid continuum limit and are therefore quite noisy. The effect of heavy pion masses in the lattice is also expected to be more distorting for $\chi_{top}$ than for other observables, due to its mass dependence commented above, and actually there is a high sensitivity to $m_\pi$ in the  results   in \cite{Astrakhantsev:2019wnp}. At most, one can infer from those results a growing tendency   $\chi_{top}(\mu_5)$ for large $\mu_5$, which is not so clear for lower values. 
  
\begin{figure}[h]
 \centerline{ \includegraphics[width=0.65\textwidth]{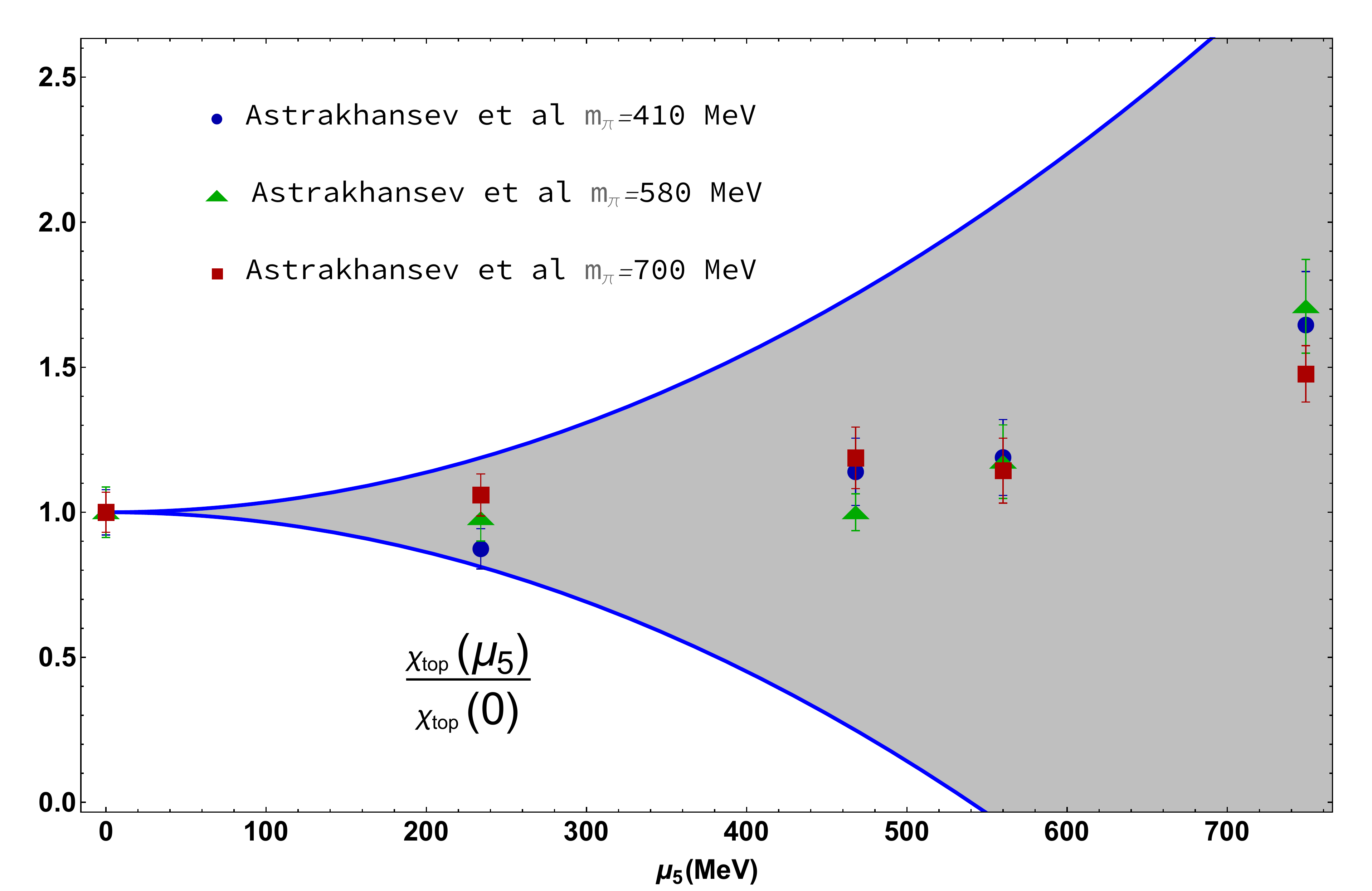}}
  \caption{$\mu_5$ dependence of the topological susceptibility chiral transition temperature extracted from the vanishing of the quark condensate. We show the uncertainty bands corresponding to $\vert \kappa_3 \vert\leq \frac{1}{16\pi^2}$.  We include 
  the lattice points from  \cite{Astrakhantsev:2019wnp} for $a=0.0856$.}
  \label{fig:chitopband}
\end{figure}

  Nevertheless, the previous uncertainties reduce considerably by considering the ratio $\chi_{top} (\mu_5)/\chi_{top}(0)$ for which the  $M^2$ dependence is expected to cancel out, as can be seen in Fig.\ref{fig:chitopband}, where the lattice data for different masses are compatible within errors. Also, the band of natural values for $\kappa_3$ covers widely the lattice points, so it is meaningful to  fit the lowest  data points with the ChPT curve, which at this order is given by

\begin{equation}
\frac{\chi_{top}(\mu_5)}{\chi_{top}(0)}=1+4\frac{\kappa_3 \mu_5^2}{F^2} + \Od(1/F^4)
\label{chitopchpt}
\end{equation}
for $m_u=m_d$, where we have used that $\chi_{top}(0)=M^2 F^2/4 + \Od(F^0)$ in $SU(2)$ \cite{Mao:2009sy,Guo:2015oxa,Nicola:2019ohb}.

\begin{figure}[h]
 \centerline{ \includegraphics[width=0.45\textwidth]{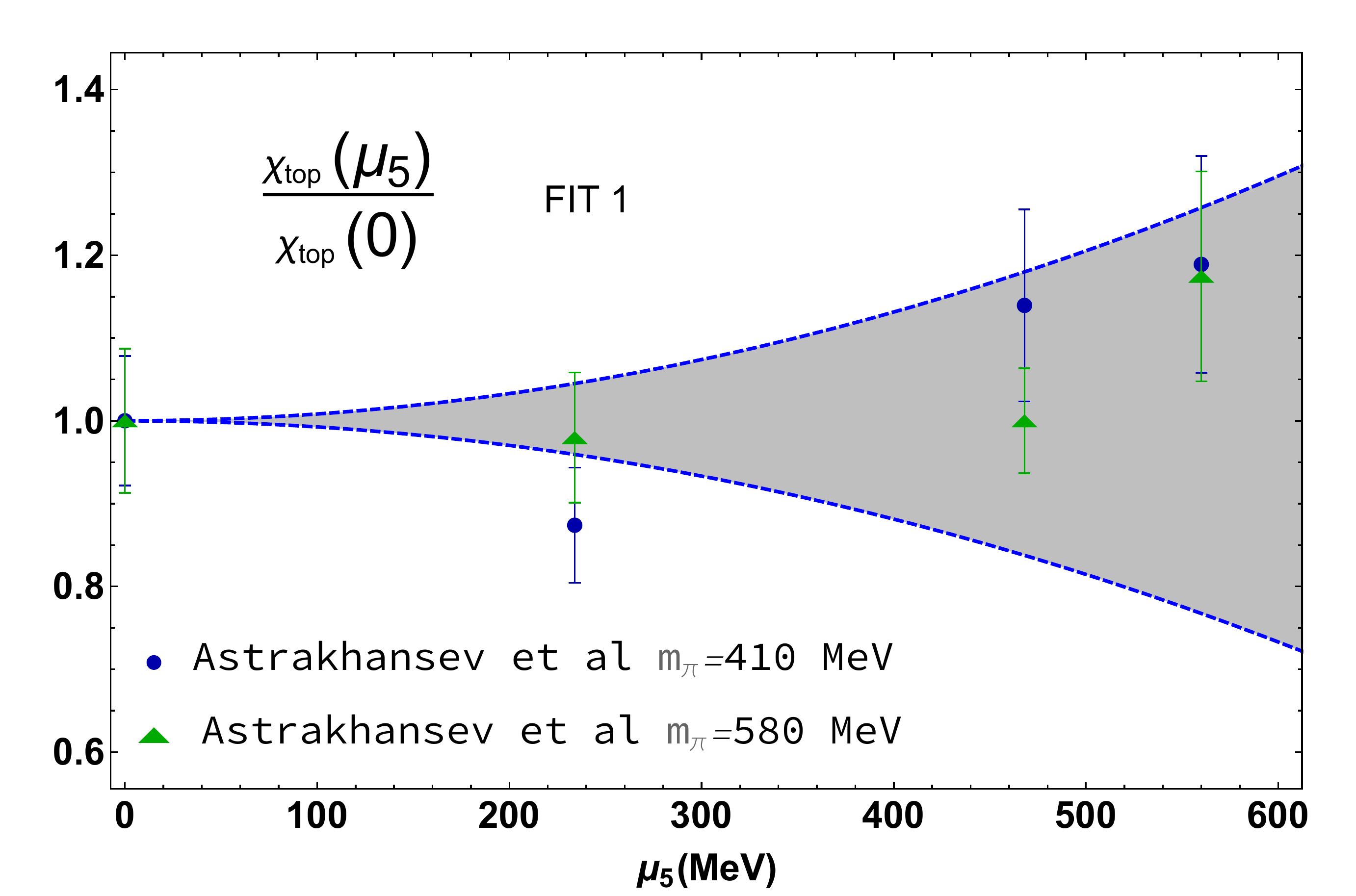}
  \includegraphics[width=0.45\textwidth]{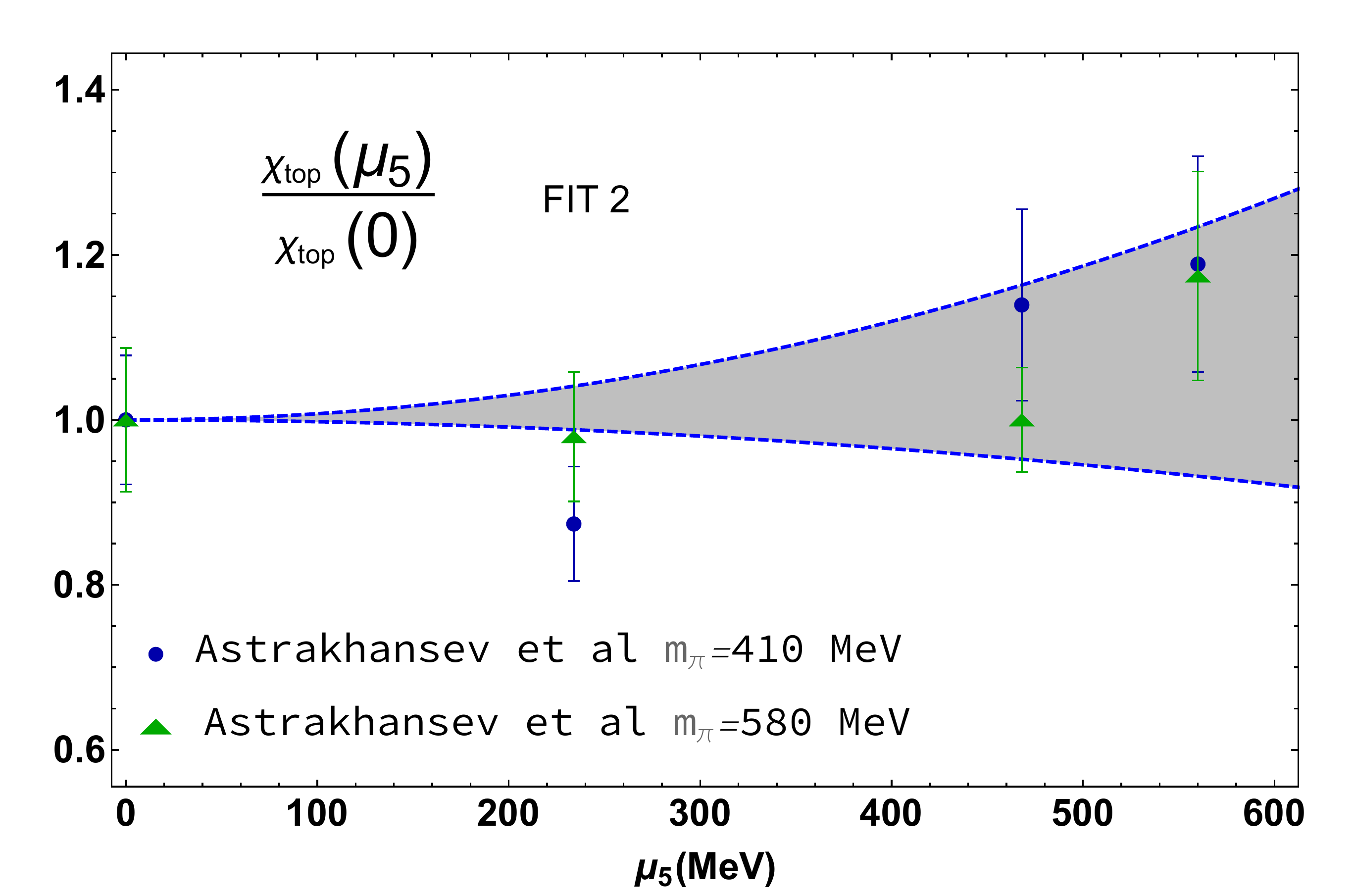}}
  \caption{Fits of $\chi_{top}(\mu_5)/\chi_{top}(0)$ using the ChPT  expression \eqref{chitopchpt}.   The lattice points used for the fit are those in \cite{Astrakhantsev:2019wnp}.}
  \label{fig:chitopfits}
\end{figure}

\begin{table}[h]
\begin{tabular}{||l|c|c|c|c||}  \hhline{|=|=|=|=|=|}
	FIT& $\kappa_3 \times 10^3$& $R^2$ &$\chi^2$/dof& \# points $\mu_5\neq 0$\ \\ \hline
	Fit 1 & 0.1 $\pm$ 1.4& 0.99&1.20&2 ($m_\pi=410$ MeV)+2 ($m_\pi=580$ MeV) \\ 
	Fit 2 & 0.5 $\pm$ 0.9& 0.99   &1.13 &3 ($m_\pi=410$ MeV)+3 ($m_\pi=580$ MeV) 
	\\ \hhline{|=|=|=|=|=|}	\end{tabular}
\caption{Numerical values of the parameters corresponding to the fits in Fig.\ref{fig:chitopfits}. The last column indicates the number of lower $\mu_5$ points considered}
\label{tab:chitopfits}
\end{table}

The results of the best fits are showed in Fig.\ref{fig:chitopfits} and Table \ref{tab:chitopfits}, where we have selected two sets for the lowest masses in \cite{Astrakhantsev:2019wnp}. The results for $\kappa_3$ are compatible with zero but the error bands corresponding to the 95\% confidence level of the fits are much narrower than the natural values showed in Fig. \ref{fig:chitopband}.

\subsection{Pressure and speed of sound}

It is  interesting to explore the consequences of the $\mu_5$ corrections and the new LEC involved, as far as other thermodynamical quantities are concerned. The thermodynamic pressure $P$, the entropy density $s$, the specific heat $c_v$,  and the speed of sound squared $c_s^2$   can be obtained from the energy density \eqref{z} as customary. In the infinite volume limit,

\begin{eqnarray}
P(T,\mu_5)&=&\lim_{V\rightarrow\infty} \left[ \epsilon(0,\mu_5)-\epsilon(T,\mu_5) \right], \nonumber\\
s(T,\mu_5)&=&\frac{\partial P(T,\mu_5)}{\partial T}, \quad c_v(T,\mu_5)=T\frac{\partial s}{\partial T}, \quad c_s^2(T,\mu_5)=\frac{\partial P}{\partial \epsilon} =\frac{s}{c_v}
\label{pressureandcsdef}
\end{eqnarray}

From our expressions from the energy density in section \eqref{sec:energyden}, we obtain for the pressure

\begin{eqnarray}
P(T,\mu_5)&=&\frac{3}{2} g_0(M,T)\left(1+6\kappa_2\frac{\mu_5^2}{F^2}\right)-\frac{3M^2}{8F^2}\left\{\left[g_1(M,T)\right]^2
\right. \nonumber \\
&+&\left.
8g_1(M,T)\left[M^2\left(l_3^r(\mu)+\frac{1}{64\pi^2}\log\frac{M^2}{\mu^2}\right)
-2\mu_5^2\kappa_b\right]
\right\}
+\Od\left(\frac{1}{F^4}\right)
\label{pressure}
\end{eqnarray}
which in the chiral limit reduces to

\begin{equation}
\left.P(T,\mu_5)\right\vert_{M=0}=\frac{\pi^2 T^4}{15}\left( 1+ 6\kappa_2\frac{\mu_5^2}{F^2} \right) 
\label{pressurechiral}
\end{equation}

\begin{figure}[h]
 \centerline{ \includegraphics[width=0.45\textwidth]{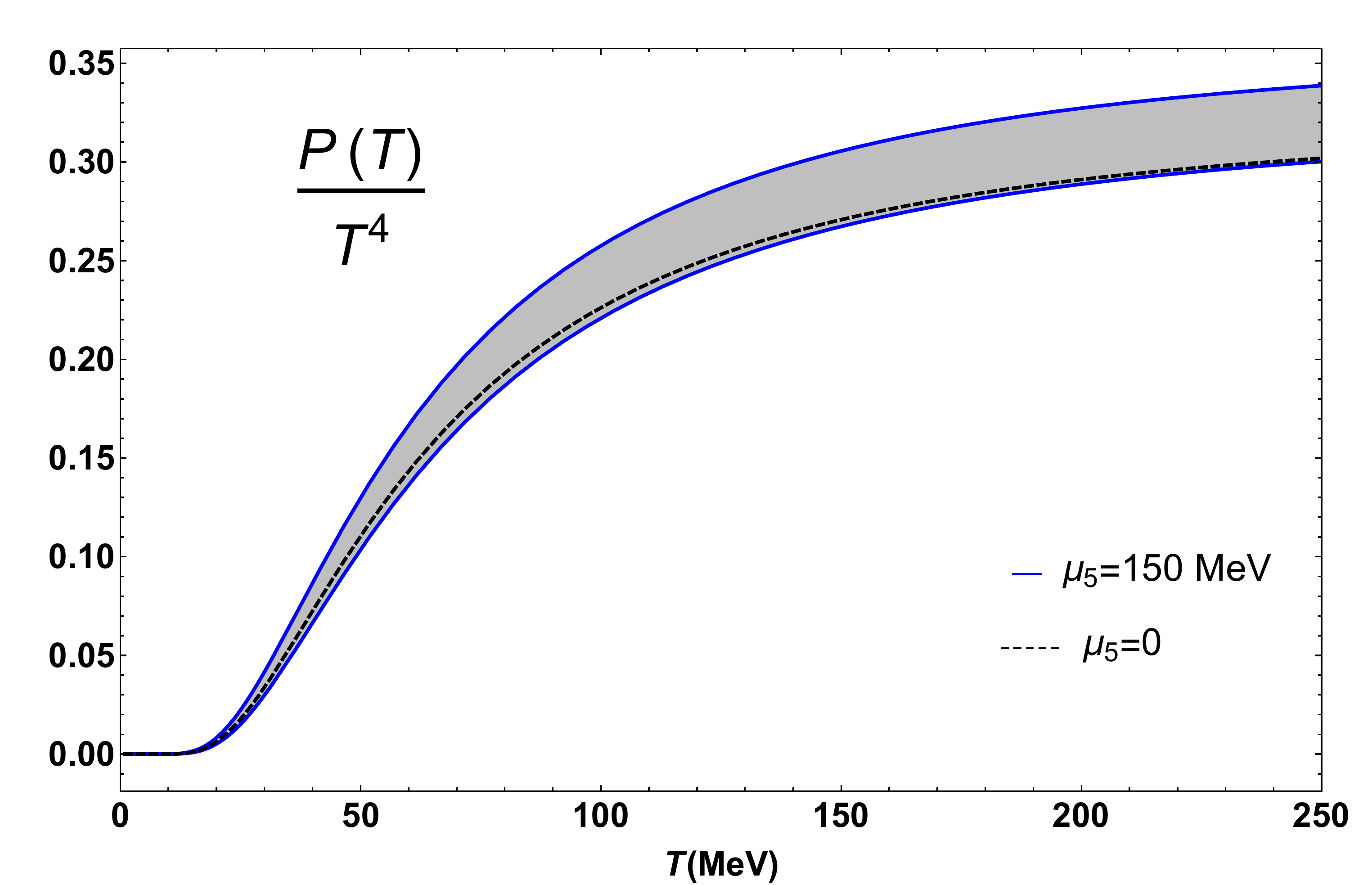}
  \includegraphics[width=0.45\textwidth]{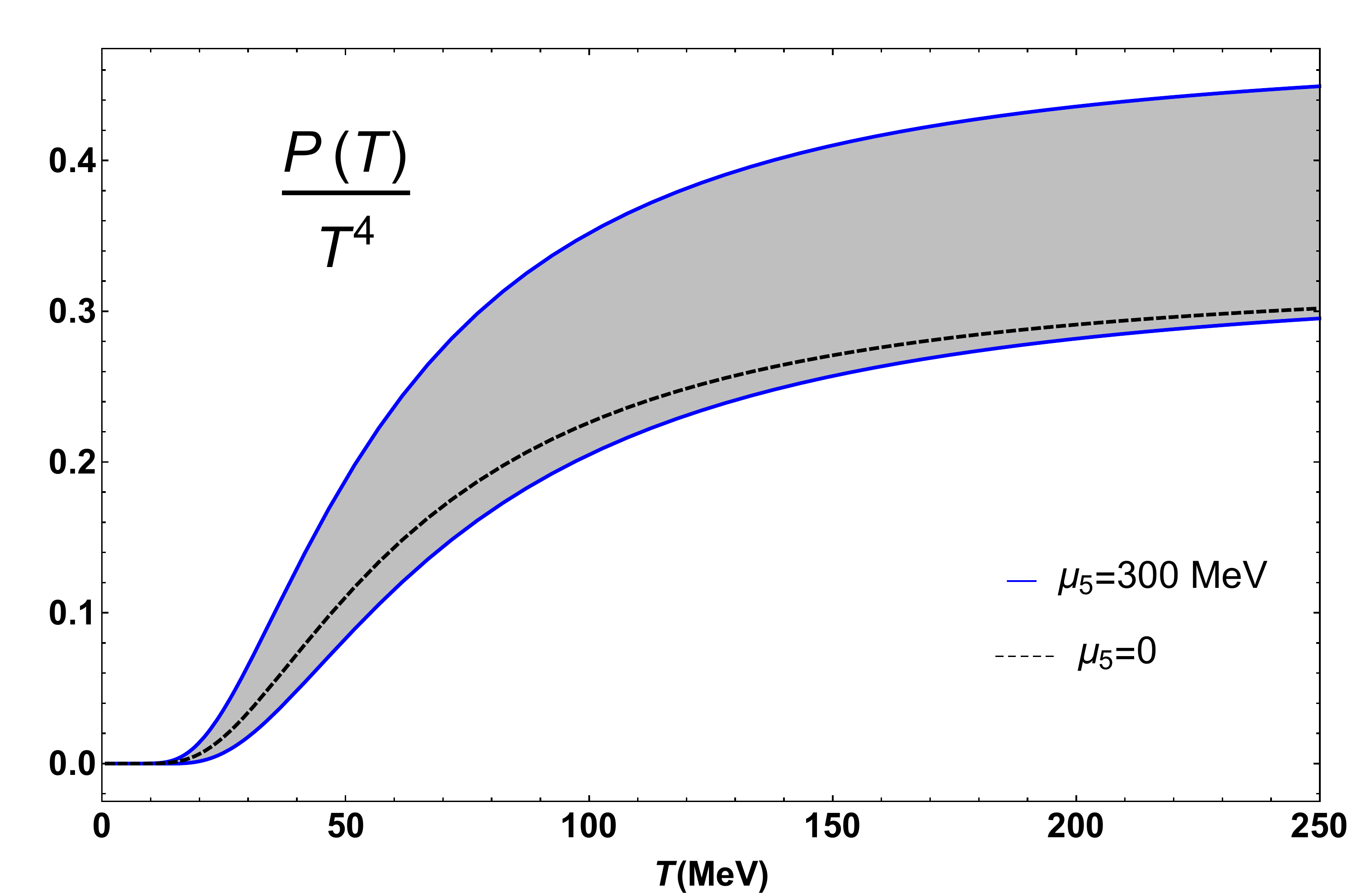}}
  \caption{Pressure for $\mu_5=150$ MeV and $\mu_5=300$ MeV compared to the $\mu_5=0$ case. The uncertainty bands correspond  to $0\leq  \kappa_2 \leq \frac{1}{16\pi^2}, \vert \kappa_b \vert\leq \frac{1}{16\pi^2}$.}
  \label{fig:pressure}
\end{figure}

Thus, the $\mu_5$ corrections to the pressure are parametrized by  $\kappa_2$ and the combination $\kappa_b$ in \eqref{kappaab}, only the first one surviving in the chiral limit, which corresponds to the ultrarelativistic free pion gas corrected by the $\kappa_2$ term. In Figure \ref{fig:pressure} we represent the pressure for two reference values of $\mu_5=150$ MeV and $\mu_5=300$ MeV for which our approach can be considered valid, with the bands corresponding to natural values for $\kappa_2,\kappa_b$, keeping $\kappa_2>0$ as explained in section \ref{sec:piondr}. The main source of uncertainty in the pressure comes actually from the $\kappa_2$ term, the result remaining almost unchanged if using for instance the uncertainty for $\kappa_b$ given in Table \ref{tab:Tcfits}.

\begin{figure}[h]
 \centerline{ \includegraphics[width=0.45\textwidth]{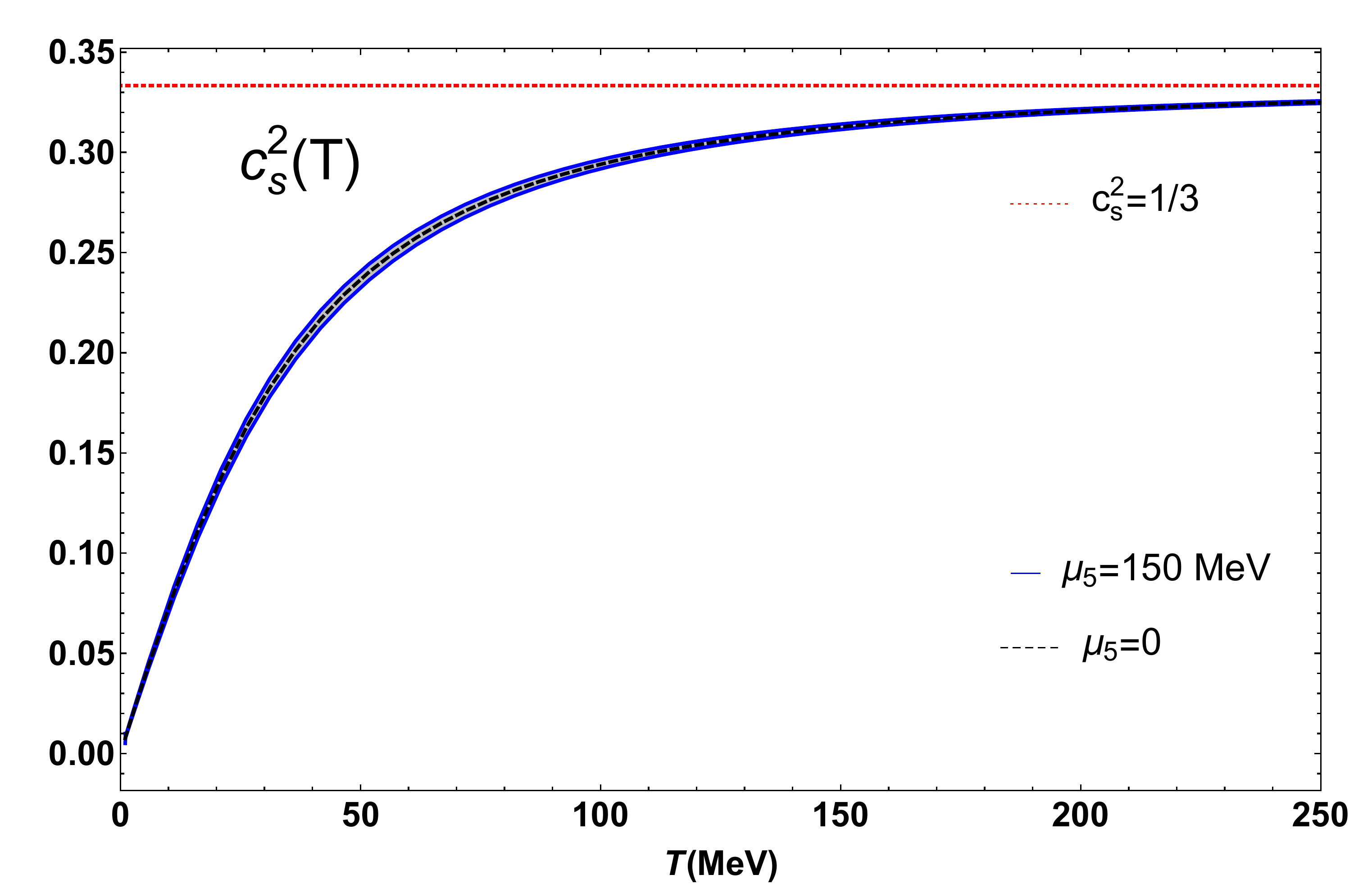}
  \includegraphics[width=0.45\textwidth]{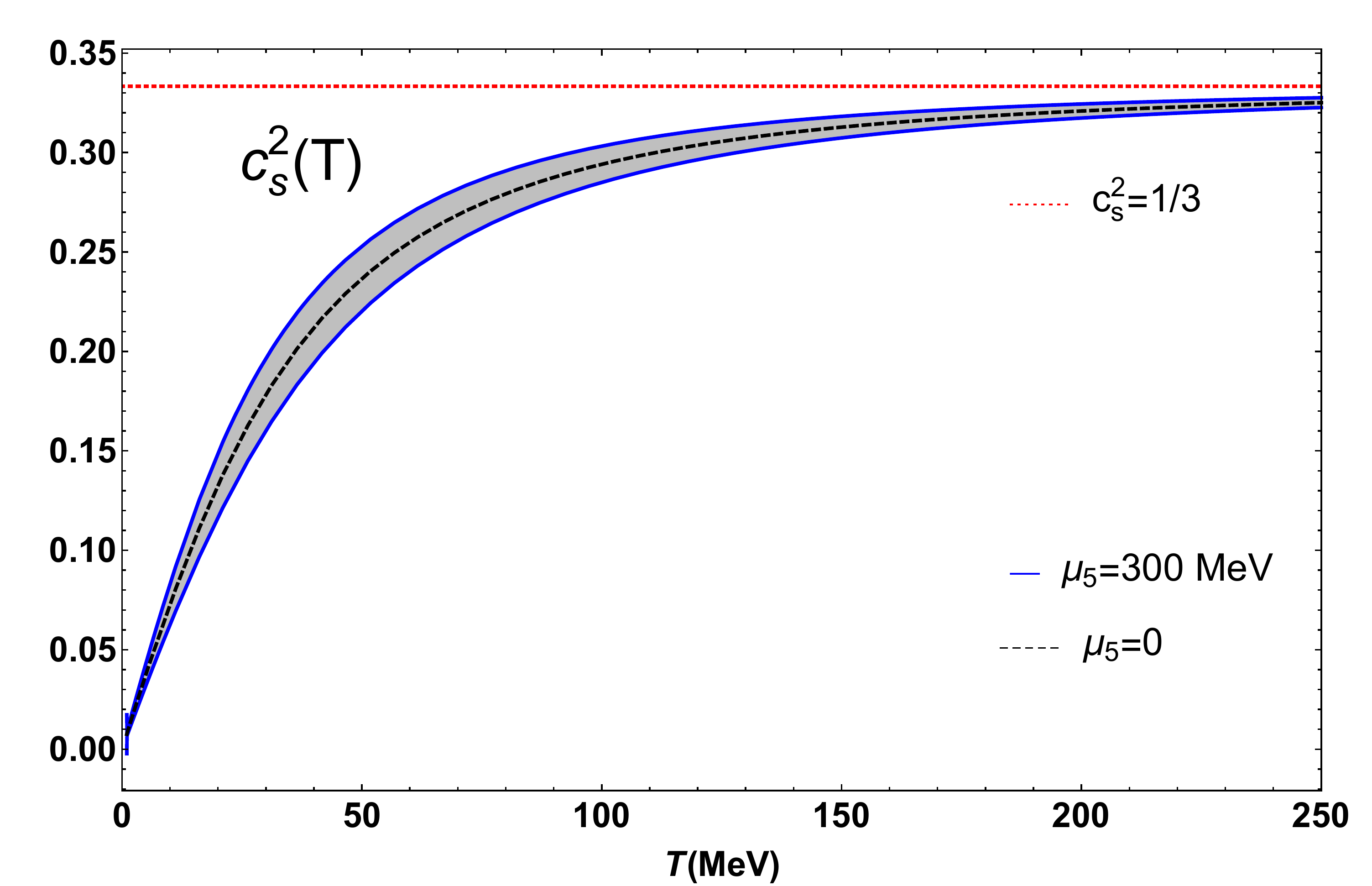}}
  \caption{Speed of sound squared for $\mu_5=150$ MeV and $\mu_5=300$ MeV compared to the $\mu_5=0$ case. The uncertainty bands correspond  to $\vert \kappa_b \vert\leq \frac{1}{16\pi^2}$.}
  \label{fig:cs}
\end{figure}

It is particularly interesting to study the speed of sound and whether the $\mu_5$ corrections may affect the physical limitations for this quantity. The analytic result for $M\neq 0$ is given by

\begin{eqnarray}
c_s^2(T,\mu_5)&=&\frac{1}{T g_0''(M,T)}\left\{ g_0'(M,T) +\frac{M^2}{2F^2 g_0''(M,T)}\left[\left(g'_1(M,T)\right)^2 g_0'(M,T)\right.\right.\nonumber\\
&+&\left.\left.\left(g_1(M,T)+4M^2\left(l_3^r(\mu)+\frac{1}{64\pi^2}\log\frac{M^2}{\mu^2}\right)-8\kappa_b\mu_5^2\right)\left(g_0'(M,T)g_1''(M,T)-g_1'(M,T)g_0''(M,T)\right)
\right]
\right\} \nonumber\\
&+&\Od\left(\frac{1}{F^4}\right)
\label{csq}
\end{eqnarray}
where $g_i'(M,T)$ and $g_i''(M,T)$ denote derivatives with respect to $T$. 
Note that the $\kappa_2$ contribution cancels in $c_s^2$ which then depends only on $\mu_5$ upon the $\kappa_b$ combination. In addition, at this order in the chiral limit one just gets $c_s^2\rightarrow 1/3$, i.e, the ultrarelativistic limit of a free boson gas, which is meant to be reached asymptotically as the temperature increases, i.e., for $T\gg M$.  This is clearly seen  in Figure \ref{fig:cs} where $c_s^2$ remains below  $1/3$ with the $\mu_5$ corrections included. The uncertainty band for $\kappa_b$ actually narrows as $T$ increases, consistent with the chiral limit being $\mu_5$-independent for this quantity at this order. Therefore, having no lattice results available to compare with, the analysis of pressure and the speed of sound poses no extra requirements on the $\kappa_i$ LEC.

 \section{Conclusions}
 
 In this work we have analyzed the effective chiral lagrangian for nonzero chiral imbalance for two light flavours, through its dependence with the axial chemical potential $\mu_5$. Our analysis provides a consistent framework for the behaviour at low and moderate values of $\mu_5$, which would be useful for physical systems where local parity breaking is at work,  as in relativistic heavy ion collisions.   Thus, we have constructed the most general lagrangian up to fourth order,  following the technique of external sources extended to include their singlet components. We have also explored the main phenomenological consequences, paying special attention to the comparison with existing lattice results. 
 
In  the lagrangian construction,  two different types of operators arise: those coming from the covariant derivative with a singlet axial field $a_\mu^0$  incorporated, which are proportional to standard low-energy constants, and new terms allowed by the symmetry in the presence of $a_\mu^0$, carrying new LEC.   The second-order lagrangian ${\cal L}_2$ only receives a constant (field independent) $\mu_5$-dependent contribution, which affects the free energy. It contains a new low-energy constant $\kappa_0$. At fourth order, there are two derivative-like terms in ${\cal L}_4$, one of them breaking Lorentz covariance (broken by the choice of $a_\mu^0$), a mass term and a constant term. All of them are multiplied by  combinations of standard and new LEC, giving rise to four undetermined constants $\kappa_{1,\dots,4}$, whose renormalization ensures that the $\mu_5$-dependent corrections to observables are finite. 

Regarding the phenomenological consequences, we have analyzed several observables and the dependence of their $\mu_5$ corrections with the $\kappa_i$ constants. The main results  are the following:

\begin{itemize}

\item The pion dispersion relation is modified through a reduction in the pion velocity due to the Lorentz breaking effect. The same effect implies that the screening and pole pion masses become different, both receiving $\mu_5$ corrections. The LEC involved are $\kappa_2$ for the pion velocity, which should be negative to ensure that pions do not become tachyonic, and  both $\kappa_2$ 
 and $\kappa_1-\kappa_3$ for the masses. At present, there are no available lattice data to confront here, the screening mass being the most feasible one, which we leave here as a suggestion for future lattice analysis. 

\item  We have calculated the $\mu_5$ corrections to the vacuum energy density up to sixth order, which includes  three new LEC coming from the sixth-order lagrangian. We have shown that the energy density is finite and scale independent once the LEC renormalization is properly accounted for. From the energy density, several observables can be extracted, such as the quark condensate, the scalar susceptibility and the chiral charge density.

\item Chiral restoration properties at finite temperature have been  explored through the light quark condensate and the scalar susceptibility. The transition temperature approximated from the vanishing of the quark condensate turns out to behave quite according to lattice analysis when the ratio $T_c(\mu_5)/T_c(0)$ is considered in the chiral limit. Actually, this allows to provide a rather trustable numerical value for the combination $\kappa_a=2\kappa_1-\kappa_2$ by fitting lattice data. That ratio is pretty insensitive to mass corrections. Actually, including those corrections in the fit gives a worse determined numerical value for the combination $\kappa_b=\kappa_1+\kappa_2-\kappa_3$, appearing for $M\neq 0$.  The analysis of the scalar susceptibilty $\chi_S$ shows that for low $\mu_5$ it is controlled by the same combination $\kappa_1-\kappa_3$ appearing in the screening mass. A positive sign for that combination would be consistent with the lattice observation  of $\chi_S$ decreasing with $\mu_5$ below the transition. 

\item The chiral charge density $\rho_5$ follows essentially a linear behaviour with $\mu_5$, consistently with lattice data. The constant $\kappa_0$ appearing in the linear term can actually be well fixed by fitting the lattice points with the chiral limit ChPT expression, since lattice data for $N_f=2$ are almost insensitive to the pion mass.  Thermal corrections are small here and reduce the size of the linear term in the chiral limit. The fourth-order constant $\kappa_4$ and the sixth-order one $\gamma_0$ enter in the cubic and fifth-order terms respectively, and can also be reasonably determined. This observed behaviour of $\rho_5$ in the lattice is at odds with the possibility of a $\mu_5\neq 0$ minimum for the free energy for low and moderate values of $\mu_5$. 

\item The topological susceptibility $\chi_{top}$ dependence with $\mu_5$ has also been determined recently in the lattice, although it is subject to more uncertainties than the previously considered observables. Our ChPT analysis  predicts that the lowest order corrections to $\chi_{top}$ are of order $\mu_5^2$ controlled by the $\kappa_3$ constant. The pion mass dependence uncertainty of lattice points is  reduced by taking the ratio  $\chi_{top}(\mu_5)/\chi_{top}(0)$, which allows to obtain a decent determination of $\kappa_3$, albeit with larger errors than other combinations. 

\item The pressure and the speed of sound are also affected by $\mu_5$ corrections, which at the order considered depend on $\kappa_2$ and $\kappa_b$ for the pressure, while for the speed of sound the $\kappa_2$ dependence disapears. The ultrarelativistic limit $T\gg M$ is reached for both quantities as temperature increases and corresponds to the chiral limit. In the case of the speed of sound, that limit is the $1/3$ value independent of $\mu_5$ which is not violated by the $\mu_5$ corrections at nonzero $M$. 

\end{itemize}
  
From the previous analysis, a consistent picture with lattice data emerges, allowing to determine some LEC combinations with acceptable precision.  Taking the average values of the $\kappa_a$ and $\kappa_3$ fits performed here in sections \ref{sec:condensate} and \ref{sec:topsus}, one gets $\kappa_1\simeq 0.8\times 10^{-3}$, $\kappa_2\simeq -0.5 \times 10^{-3}$, $\kappa_3 \simeq 0.3 \times 10^{-3}$ although with uncertainties of order $10^{-3}$, i.e., larger than for particular combinations such as $\kappa_a$, inherited from the uncertainty in $\kappa_b$. Nevertheless, it is remarkable to observe that those mean values obey the  expected sign conditions discussed above, namely $\kappa_2<0$ and $\kappa_1-\kappa_3 >0$. Apart from describing the observed lattice trends for the observables mentioned above, our analysis points to decreasing pion mass and increasing pion decay constant consistently with recent model analyses.   Further lattice observables, such as the screening masses, or improving the precision over existing determinations, would certainly help to narrow this picture.

Summarizing, our present study provides a solid setup for the analysis of chirally imbalanced matter for two light flavours, at low and moderate values of $\mu_5$, typically $\mu_5 \lsim$ 500 MeV. A rigorous construction of the efective lagrangian has been developed and the main physical effects have been analyzed and compared to existing lattice data, which allows for a first determination of the new low-energy constants involved. Thus, within its applicability range, the present analysis is meant to provide a  useful  benchmark  for  model and lattice analysis.

 \section*{Acknowledgments}
 We are very grateful to   A.A.Andrianov, V.A. Andrianov  and J.Urrestarazu for useful discussions and comments. Work partially supported by  research contracts FPA2016-75654-C2-2-P, FPA2016-76005-C2-1-P, MDM-2014-0309  (Ministerio de Econom\'{\i}a y Competitividad) and 2017SGR929 (Generalitat de Catalunya). This work has also received funding from the European Union Horizon 2020 research and innovation programme under grant agreement No 824093. A. V-R acknowledges support from a fellowship of the UCM predoctoral program.

\appendix

\section{$\Od(p^3)$  operators}
\label{sec:op3}

Here, we list all the possible $\Od(p^3)$ terms for arbitrary $Q_L$ and $Q_R$, compatible with the symmetries. Those terms can be of the following types:

\begin{itemize}

\item {\em $\chi Q$-like terms:}

\begin{align}
\tr\left[Q_L(U^\dagger \chi +\chi^\dagger U)+Q_R(U\chi^\dagger+\chi U^\dagger)\right]\nonumber\\
\tr\left[\chi^\dagger U+\chi U^\dagger\right]\tr[Q_L+Q_R]
\end{align}

\item {\em $QQQ$ terms:}
\begin{align}
\tr[Q_R^2 U Q_L U^\dagger+Q_L^2 U^\dagger Q_R U] \nonumber\\
\tr[Q_R^3+Q_L^3] \nonumber\\
\tr[Q_L+Q_R]\tr\left[Q_R U Q_L U^\dagger  \right] \nonumber\\
\tr[Q_L+Q_R]\tr[Q_R^2+Q_L^2]\nonumber\\
\tr(Q_L-Q_R)\tr (Q_L^2-Q_R^3)\nonumber\\
\tr^2[Q_L]\tr(Q_R)+ \tr^2[Q_R]\tr(Q_L) \nonumber\\
\tr^3[Q_L]+\tr^3[Q_R]
\end{align}

\item{\em $Qdd$ terms} 
\begin{align}
\tr[Q_L+Q_R]\tr[d_\mu U^\dagger d^\mu U] \nonumber\\
\tr[Q_Ld_\mu U^\dagger d^\mu U+Q_Rd_\mu U d^\mu U^\dagger]\nonumber\\
\tr[Q_L+Q_R]\tr(U^\dagger d_\mu U)\tr(U^\dagger d^\mu U)
\end{align}

\item{\em $cQd$ terms:} 
Terms of the form $\tr[(c_{\mu L} Q_L) U^\dagger d^\mu U]$, $\tr[(c_{\mu R}) Q_R U d^\mu U^\dagger]$ are not possible since there is no combination of them which can be made $P$ and $C$ invariant (using $Ud_\mu U^\dagger=-d_\mu U U^\dagger$ ) and so on for  terms  $\tr[c_{\mu {L,R}}  Q_{L,R}] \tr[U^\dagger d^\mu U]$.

\end{itemize}

\section{ $O(p^4)$ operators}
\label{app:op4}

\subsection{ $SU(2)$ operator identities}
\label{app:iden}

	For arbitrary two-dimensional matrices $A_1$, $A_2$, $A_3$, Cayley-Hamilton theorem implies \cite{Scherer:2002tk}:

	\begin{equation}
	\tr (A_3\{A_1,A_2\})=\tr(A_1)\tr(A_2 A_3)+\tr(A_2)\tr(A_1 A_3)+ \tr(A_3)\tr(A_1 A_2) - \tr(A_1)\tr(A_2)\tr(A_3)
	\label{traceidsu2}
	\end{equation}
	
	Using \eqref{traceidsu2} we can eliminate single traces of operators in terms of double or triple traces. In particular, the following identities hold:

	\begin{eqnarray}
	\tr(d_\mu U^\dagger d^\mu U d_\nu U^\dagger d^\nu U)&=&-\tr(U^\dagger d_\mu U) \tr(U^\dagger d^\mu U d^\nu U^\dagger d_\nu U)+\frac{1}{2}\left[ 
	\tr(U^\dagger d_\mu U)\tr(U^\dagger d^\mu U)\tr(d^\nu U^\dagger d_\nu U) \right. \nonumber\\
	&+&\left. \tr^2(d_\mu U^\dagger d^\mu U)
	\right]
	\label{trace4d1}
	\end{eqnarray}

	\begin{eqnarray}
	\tr(d_\mu U^\dagger d_\nu U d^\mu U^\dagger d^\nu U)&=&-\tr(U^\dagger d_\mu U) \tr(U^\dagger d^\mu U d^\nu U^\dagger d_\nu U) +
	\tr(U^\dagger d_\mu U) \tr(U^\dagger d_\nu U) \tr(d^\mu U^\dagger d^\nu U) \nonumber\\
	&-&\frac{1}{2}\left[ 
	\tr(U^\dagger d_\mu U) \tr(U^\dagger d^\mu U) \tr(d^\nu U^\dagger d_\nu U) + \tr^2(d_\mu U^\dagger d^\mu U) 
	\right]
	+\tr(d_\mu U^\dagger d_\nu U) \tr( d^\mu U^\dagger d^\nu U)\nonumber \\
	\label{trace4d2}
	\end{eqnarray}

	\begin{eqnarray}
	 2 \hspace{0.1cm} \tr(U^\dagger d^\mu U d^\nu U^\dagger d_\nu U)&=&2\hspace{0.1cm}\tr(U^\dagger d_\nu U)\tr(d_\mu U^\dagger d^\nu U) + \tr(U^\dagger d_\mu U)\tr(d_\nu U^\dagger d^\nu U)\nonumber\\
	 &+&\tr(U^\dagger d_\mu U)	\tr(U^\dagger d_\nu U)\tr(U^\dagger d^\nu U)
	 \label{trace4d3}
	 \end{eqnarray}

	\begin{eqnarray}
	\tr \left[d_\mu U^\dagger d^\mu U\left( \chi^\dagger U + U^\dagger \chi \right)\right]&=& \frac{1}{2}\tr(d_\mu U^\dagger d^\mu U) 
	\tr( \chi^\dagger U + U^\dagger \chi) - \tr(U^\dagger d_\mu U)\tr\left[U^\dagger d_\mu U \left( \chi^\dagger U + U^\dagger \chi \right)\right] \nonumber\\
	&+&\frac{1}{2}\tr(U^\dagger d_\mu U) \tr(U^\dagger d^\mu U)\tr( \chi^\dagger U + U^\dagger \chi)
	\label{trace2dchi1}
	\end{eqnarray}
	
	\begin{equation}
	\begin{split}
	&\text{tr}\left(d^{\mu}U^{\dagger}d_{\mu}UQ_L^2+d^{\mu}Ud_{\mu}U^{\dagger}Q_R^2\right)=\frac{1}{2}\text{tr}\left(d^{\mu}U^{\dagger}d_{\mu}U\right)\text{tr}\left(Q_R^2+Q_L^2\right)\\
	&-\text{tr}\left(d^{\mu}U^{\dagger}U\right)\text{ tr}\left(Ud_{\mu}U^{\dagger}Q_R^2-U^{\dagger}d_{\mu}UQ_L^2\right)-\frac{1}{2}\text{tr}\left(d^{\mu}U^{\dagger}U\right)\text{tr}\left(U^{\dagger}d_{\mu}U\right)\text{ tr}\left(Q_R^2+Q_L^2\right)
	\end{split}
	\label{trace2d2Q1}
	\end{equation}
	
	\begin{equation}
	\begin{split}
	&\text{tr}\left(d^{\mu}U^{\dagger}d_{\mu}UQ_L+d^{\mu}Ud_{\mu}U^{\dagger}Q_R\right)=\frac{1}{2}\text{tr}\left(d^{\mu}U^{\dagger}d_{\mu}U\right)\text{tr}\left(Q_R+Q_L\right)\\
	&-\text{tr}\left(d^{\mu}U^{\dagger}U\right)\text{ tr}\left(Ud_{\mu}U^{\dagger}Q_R-U^{\dagger}d_{\mu}UQ_L\right)-\frac{1}{2}\text{tr}\left(d^{\mu}U^{\dagger}U\right)\text{tr}\left(U^{\dagger}d_{\mu}U\right)\text{ tr}\left(Q_R+Q_L\right)
	\end{split}
	\end{equation}
	
	\begin{equation}
	\begin{split}
	&\text{tr}\left(d^{\mu}U^{\dagger}d_{\mu}UQ_LU^{\dagger}Q_RU+d_{\mu}Ud^{\mu}U^{\dagger}Q_RUQ_LU^{\dagger}\right)=\\
	&\text{tr}\left(d^{\mu}U^{\dagger}d_{\mu}U\right)\text{tr}\left(Q_LU^{\dagger}Q_RU\right)+\text{tr}\left(d^{\mu}U^{\dagger}U\right)\text{tr}\left(d_{\mu}UQ_LU^{\dagger}Q_R-d_{\mu}U^{\dagger}Q_RUQ_L\right)\\
	&-\text{tr}\left(d^{\mu}U^{\dagger}U\right)\text{tr}\left(U^{\dagger}d^{\mu}U\right)\text{tr}\left(Q_RUQ_LU^{\dagger}\right)
	\end{split}
	\label{trace2d2Q2}
	\end{equation}
	
	\begin{equation}
	\begin{split}
	&2 \hspace{0.1cm}\tr (d_{\mu}UQ_LU^{\dagger}Q_R-d_{\mu}U^{\dagger}Q_RUQ_L)=\tr(U^{\dagger}d_{\mu}U)\tr(Q_LU^{\dagger}Q_RU)+\tr(d_{\mu}UU^{\dagger}Q_R)\tr(Q_L)\\
	&+\tr(U^{\dagger}d_{\mu}UQ_L)\tr(Q_R)-\tr(U^{\dagger}d_{\mu}U)\tr(Q_L)\tr(Q_R)
	\end{split}
	\end{equation}
	
	\begin{equation}
	\begin{split}
	&\text{tr}\left[\left(\chi U^{\dagger}+U\chi^{\dagger}\right)Q_{R}^2+\left(\chi^{\dagger} U+U^{\dagger}\chi\right)Q_{L}^2\right]=\frac{1}{2}\text{tr}\left(\chi^{\dagger}U+U^{\dagger}\chi\right)\text{tr}\left[\left(Q_R^2\right)+\left(Q_L^2\right)\right]\\
	&+\text{tr}\left[\left(\chi U^{\dagger}+U\chi^{\dagger}\right)Q_R\right]\text{tr}\left(Q_R\right)+\text{tr}\left[\left(\chi^{\dagger} U+U^{\dagger}\chi\right)Q_L\right]\text{tr}\left(Q_L\right)-\frac{1}{2}\left[\text{tr}^2\left(Q_L\right)+\text{tr}^2\left(Q_R\right)\right]\text{tr}\left(\chi^{\dagger}U+U^{\dagger}\chi\right)
	\end{split}
	\end{equation}

	\begin{equation}
	\begin{split}
	&\text{tr}\left[\left(\chi^{\dagger}U+U^{\dagger}\chi\right)Q_LU^{\dagger}Q_RU+\left(\chi U^{\dagger}+U\chi^{\dagger}\right)Q_RUQ_LU^{\dagger}\right]=\text{tr}\left(\chi U^{\dagger}+U\chi^{\dagger}\right)\text{tr}\left(Q_RUQ_LU^{\dagger}\right)\\
	&+\text{tr}\left[\left(\chi U^{\dagger}+U\chi^{\dagger}\right)Q_R\right]\text{tr}\left(Q_L\right)+\text{tr}\left[\left(\chi^{\dagger} U+U^{\dagger}\chi\right)Q_L\right]\text{tr}\left(Q_R\right)-\text{tr}\left(Q_R\right)\text{tr}\left(Q_L\right)\text{tr}\left(\chi U^{\dagger}+U\chi^{\dagger}\right)
	\end{split}
	\label{trace2M2Q1}
	\end{equation}

	\begin{equation}
	\begin{split}
	&\text{tr}\left[\left(Q_RUQ_LU^{\dagger}\right)^2\right]=\text{tr}^2\left(Q_RUQ_LU^{\dagger}\right)+\frac{1}{2}\left[\text{tr}\left(Q_R\right)\text{tr}\left(Q_RUQ_L^2U^{\dagger}\right)+\text{tr}\left(Q_L\right)\text{tr}\left(Q_R^2UQ_LU^{\dagger}\right)\right]\\
	&-\text{tr}\left(Q_R\right)\text{tr}\left(Q_L\right)\text{tr}\left(Q_RUQ_LU^{\dagger}\right)-\frac{1}{2}\text{ tr}\left(Q_R^2\right)\text{tr}\left(Q_L^2\right)+\frac{1}{4}\text{ tr}\left(Q_R^2\right)\text{tr}^2\left(Q_L\right)+\frac{1}{4}\text{ tr}\left(Q_L^2\right)\text{tr}^2\left(Q_R\right)
	\end{split}
	\label{trace4Q1}
	\end{equation}
	
	\begin{equation}
	\begin{split}
	&\text{tr}\left(Q_R^2UQ_L^2U^{\dagger}\right)=\frac{1}{2}\text{ tr}\left(Q_R^2\right)\text{tr}\left(Q_L^2\right)+\frac{1}{2}\left[\text{tr}\left(Q_R\right)\text{tr}\left(Q_RUQ_L^2U^{\dagger}\right)+\text{tr}\left(Q_L\right)\text{tr}\left(Q_R^2UQ_LU^{\dagger}\right)\right]\\
	&-\frac{1}{4}\text{ tr}\left(Q_R^2\right)\text{tr}^2\left(Q_L\right)-\frac{1}{4}\text{ tr}\left(Q_L^2\right)\text{tr}^2\left(Q_R\right)
	\end{split}
	\label{trace4Q2}
	\end{equation}
	
	\begin{equation}
	\begin{split}
	&\text{tr}\left(Q_R\right)\text{tr}\left(Q_RUQ_L^2U^{\dagger}\right)+\text{tr}\left(Q_L\right)\text{tr}\left(Q_R^2UQ_LU^{\dagger}\right)=2\text{ tr}\left(Q_R\right)\text{tr}\left(Q_L\right)\text{tr}\left(Q_RUQ_LU^{\dagger}\right)\\
	&-\text{tr}^2\left(Q_L\right)\text{tr}^2\left(Q_R\right)+\frac{1}{2}\text{ tr}\left(Q_R^2\right)\text{tr}^2\left(Q_L\right)+\frac{1}{2}\text{ tr}\left(Q_L^2\right)\text{tr}^2\left(Q_R\right)
	\end{split}
	\label{trace4Q3}
	\end{equation}
	
	\begin{equation}
	\tr(Q_R^3)=\dfrac{3}{2}\tr(Q_R)\tr(Q_R^2)-\dfrac{1}{2}\tr^3(Q_R)
	\label{trace4Q4}
	\end{equation}
	
	\begin{equation}
	\tr(Q_L^3)=\dfrac{3}{2}\tr(Q_L)\tr(Q_L^2)-\dfrac{1}{2}\tr^3(Q_L)
	\label{trace4Q5}
	\end{equation}
	
	\begin{equation}
	\tr(Q_R^4)=\dfrac{3}{2}\tr^2(Q_R)\tr(Q_R^2)-\tr^4(Q_R)+\dfrac{1}{2}     \tr^2(Q_R^2)
	\label{trace4Q6}
	\end{equation}

	\begin{equation}
	\tr(Q_L^4)=\dfrac{3}{2}\tr^2(Q_L)\tr(Q_L^2)-\tr^4(Q_L)+\dfrac{1}{2}     \tr^2(Q_L^2)
	\label{trace4Q7}
	\end{equation}

Two useful additional relations are
	
	\begin{equation}
\begin{split}
&\text{ tr}\left[d_{\mu} \left(d^{\mu}U^{\dagger}U Q_L U^{\dagger}Q_R U\right)\right]=\text{tr}\left(d_{\mu}d^{\mu}U^{\dagger}U Q_LU^{\dagger}Q_RU\right)+2\text{ tr}\left(d^{\mu}U^{\dagger}d_{\mu}UQ_LU^{\dagger}Q_RU\right)\\
&-\text{tr}\left(d^{\mu}UQ_Ld_{\mu}U^{\dagger}Q_R\right)-\text{tr }\left(d^{\mu}U\partial_{\mu}Q_LU^{\dagger}Q_R\right)-\text{tr }\left(d^{\mu}UQ_LU^{\dagger}\partial_{\mu}Q_R\right)
\end{split}
	\end{equation}
	
	\begin{equation}
	\begin{split}
	&\text{tr}\left[d_{\mu}\left(U^{\dagger}d^{\mu}UQ_LU^{\dagger}Q_RU\right)\right]=\\
	&\text{tr}\left(U^{\dagger}d^{\mu}d_{\mu}UQ_LU^{\dagger}Q_RU\right)+\text{tr}\left(d^{\mu}UQ_Ld_{\mu}U^{\dagger}Q_R\right)+\text{tr }\left(d^{\mu}U\partial_{\mu}Q_LU^{\dagger}Q_R\right)+\text{tr }\left(d^{\mu}UQ_LU^{\dagger}\partial_{\mu}Q_R\right)
	\end{split}
	\end{equation}
		
	Using that and the equation of motion we obtain

	\begin{equation}
	\begin{split}
	\text{tr}\left(d_{\mu}U^{\dagger}Q_Rd^{\mu}UQ_L\right)=&\frac{1}{2}\text{tr}\left(d_{\mu}U^{\dagger}d^{\mu}UQ_LU^{\dagger}Q_RU+d_{\mu}Ud^{\mu}U^{\dagger}Q_RUQ_LU^{\dagger}\right)\\
	&+\frac{1}{4}\text{tr}\left[\left(\chi^{\dagger}U-U^{\dagger}\chi\right)Q_LU^{\dagger}Q_RU+\left(\chi U^{\dagger}-U\chi^{\dagger}\right)Q_RUQ_LU^{\dagger}\right]\\
	&-2Z\text{tr}\left(Q_R^2UQ_L^2U^{\dagger}\right)+2 Z\text{tr}\left[\left(Q_RUQ_LU^{\dagger}\right)^2\right]\\
	&-\frac{1}{2}\text{tr}\left[d_{\mu}U\partial^{\mu}Q_L U^{\dagger}Q_R+d_{\mu}UQ_L U^{\dagger}\partial^{\mu}Q_R\right.\\
	&\left.+d_{\mu}U^{\dagger}\partial^{\mu}Q_R UQ_L+d_{\mu}U^{\dagger}Q_R U\partial^{\mu}Q_L\right]\\
	&+\frac{1}{2}d_{\mu}\text{tr}\left[d^{\mu}UQ_LU^{\dagger}Q_R+d^{\mu}U^{\dagger}Q_RUQ_L\right]
	\end{split}
	\label{trace2d2Qadd}
	\end{equation}

\subsection{Terms with no $Q$ fields} 
\label{app:noQ}

We begin by considering four-derivative operators. The  possible terms are:

\begin{eqnarray}
\left[\tr \left(  d_\mu U^\dagger d^\mu U  \right) \right]^2 \nonumber\\
\tr \left(  d_\mu U^\dagger d_\nu U  \right) \tr \left(  d^\mu U^\dagger d\nu U  \right)   \nonumber\\
\tr \left(  d_\mu U^\dagger d^\mu U  d_\nu U^\dagger d^\nu U  \right)   \nonumber\\
\tr \left(  d_\mu U^\dagger d_\nu U  d^\mu U^\dagger d^\nu U  \right) \nonumber \\
\tr(U^\dagger d_\mu U) \tr(U^\dagger d^\mu U d^\nu U^\dagger d_\nu U) \nonumber \\
\tr(U^\dagger d_\mu U) \tr(U^\dagger d^\mu U) \tr( d^\nu U^\dagger d_\nu U) \nonumber \\
\tr(U^\dagger d_\mu U) \tr(U^\dagger d_\nu U) \tr( d^\mu U^\dagger d^\nu U) \nonumber \\
\tr(U^\dagger d_\mu U) \tr(U^\dagger d^\mu U) \tr(U^\dagger d_\nu U)  \tr(U^\dagger d^\nu U)
\label{op4termsder}
\end{eqnarray}

As customary, one can use $SU(2)$ identities to eliminate some of these operators in favor of the rest. In particular, using the identities \eqref{trace4d1}, \eqref{trace4d2} and \eqref{trace4d3} one can eliminate the third, four and fifth terms in \eqref{op4termsder}. 

As for operators including the $\chi$ field, the following terms are allowed for constant $\chi$:

\begin{eqnarray}
\tr \left[  d_\mu U^\dagger d^\mu U \left(\chi U^\dagger + U \chi^\dagger \right) \right] \nonumber \\
\tr \left(  d_\mu U^\dagger d^\mu U\right) \tr \left(\chi U^\dagger + U \chi^\dagger \right) \nonumber \\
\tr \left( U^\dagger d_\mu U\right) \tr \left[ U^\dagger d^\mu U \left(\chi U^\dagger + U \chi^\dagger \right) \right] \nonumber \\
\tr \left( U^\dagger d_\mu U\right) \tr \left( U^\dagger d^\mu U\right)  \tr \left(\chi U^\dagger + U \chi^\dagger \right)  \nonumber\\
\tr\left[ \chi^\dagger U \chi^\dagger U + U^\dagger\chi U^\dagger \chi  \right]  \nonumber\\
\tr^2\left(\chi U^\dagger + U \chi^\dagger \right) \nonumber\\
\tr^2\left(\chi U^\dagger - U \chi^\dagger \right) \nonumber\\
\tr\left(\chi^\dagger \chi\right) \nonumber \\
\re(\det\chi)
\label{op4termsdermass}
\end{eqnarray}

Using \eqref{trace2dchi1}, the first operator on \eqref{op4termsdermass} can be eliminated.

	\subsection{\hspace{0.05cm}(cQ)\hspace{0.05cm}Qd terms}
	\label{app:cQQd}
		\begin{eqnarray}
&&	\text{tr}\left(d_{\mu}U^{\dagger}\left[\left(c_R^{\mu}Q_R\right),Q_R\right]U+d_{\mu}U\left[\left(c_L^{\mu}Q_L\right),Q_L\right]U^{\dagger}\right)
	\nonumber\\
&&	\text{tr}\left[d_{\mu}U^{\dagger}Q_RU\left(c^{\mu}_LQ_L\right)+d_{\mu}UQ_LU^{\dagger}\left(c^{\mu}_RQ_R\right)\right] 
	+ \tr \left[U^\dagger Q_R d_\mu U \left( c^{\mu}_LQ_L \right)+ U Q_L d_\mu U^\dagger \left( c^{\mu}_RQ_R \right)\right]
	\nonumber\\
	&& \tr (U^\dagger d_\mu U) (\tr\left[ Q_L \left( c^{\mu}_LQ_L \right) \right] - \tr\left[ Q_R \left( c^{\mu}_RQ_R \right) \right] )	
	\nonumber \\
	 &&\tr (U^\dagger d_\mu U) (   \tr\left[  \left( c^{\mu}_LQ_L \right)  U^\dagger Q_R U \right]  -   \tr\left[  \left( c^{\mu}_RQ_R \right)  U Q_L U^\dagger \right] )
	 \nonumber\\
&&	\tr (U^\dagger d_\mu U) (   \tr(Q_L) \tr \left( c^{\mu}_LQ_L \right)  -  \tr(Q_R) \tr \left( c^{\mu}_RQ_R \right)  )
	\nonumber\\
&&	\tr (U^\dagger d_\mu U) (   \tr(Q_L) \tr \left( c^{\mu}_RQ_R \right)  -  \tr(Q_R) \tr \left( c^{\mu}_LQ_L \right)  )
\label{cQQdterms}
\end{eqnarray}

	\subsection{(cQ)\hspace{0.1cm}(cQ) terms}
	\label{app:cQcQ}
	\begin{eqnarray}
&&	\text{tr}\left[\left(c_R^{\mu}Q_R\right)U\left(c_{\mu L}Q_L\right)U^{\dagger}\right]
\nonumber\\
&&	\text{tr}\left[\left(c_R^\mu Q_R\right)\left(c_{\mu R}Q_R\right)+\left(c_L^\mu Q_L\right)\left(c_{\mu L}Q_L\right)\right]
\nonumber\\
&&	\text{tr}\left(c_R^\mu Q_R\right)\text{tr}\left(c_{\mu  R}Q_R\right) + \text{tr}\left(c_L^\mu Q_L\right)\text{tr}\left(c_{\mu L}Q_L\right)
\label{cQcQterms}
\end{eqnarray}

\section{Useful integrals in dimensional regularization}
\label{app:int}

We quote here the integrals needed for the renormalization of the vacuum energy density:

\begin{eqnarray}
\int \frac{d^{D-1} k}{(2\pi)^{D-1}} (k^2+M^2)^{-1}&=&G(x=0,T=0)=2M^2\left[\lambda + \frac{1}{32\pi^2}\log\frac{M^2}{\mu^2}\right]
\label{DR0} \\
\int \frac{d^{D-1} k}{(2\pi)^{D-1}} (k^2+M^2)^{1/2}&=&M^4\left[\lambda+ \frac{1}{32\pi^2}\log\frac{M^2}{\mu^2} - \frac{1}{64\pi^2}\right]
\label{DR1} \\
\int \frac{d^{D-1} k}{(2\pi)^{D-1}} (k^2+M^2)^{-1/2}\left(Ak^2+BM^2\right)&=&-M^4\left[(3A-4B)\left(\lambda+ \frac{1}{32\pi^2}\log\frac{M^2}{\mu^2}\right) + \frac{A}{64\pi^2} \right]
\label{DR2}
\end{eqnarray}
with $\lambda$ defined in \eqref{lambda}.

\end{document}